\documentclass[aip,reprint]{revtex4-1}
\usepackage[english]{babel}
\usepackage{mathtools}

\usepackage{tikz}

\usepackage{amssymb}
\usepackage{amsmath}
\usepackage{graphicx}

\usepackage{caption}
\usepackage{subcaption}
\usepackage{hyperref}
\usepackage{pgfplots}

\newcommand{\be}{\begin{equation}}
\newcommand{\en}{\end{equation}}

\def\bga#1\ega{\begin{gather}#1\end{gather}} % suggested in technote.tex
\def\bgas#1\egas{\begin{gather*}#1\end{gather*}}

\def\bal#1\eal{\begin{align}#1\end{align}} % suggested in technote.tex
\def\bals#1\eals{\begin{align*}#1\end{align*}}

\renewcommand{\vec}[1]{\boldsymbol{#1}}

\newcommand{\ii}{\textrm{i}}
\newcommand{\ee}{\textrm{e}}

\newcommand{\R}{\mathbb{R}}

\usepackage[colorinlistoftodos,bordercolor=orange,backgroundcolor=orange!20,linecolor=orange,textsize=scriptsize]{todonotes}
\usepackage{soul}

\newcommand{\ensem}[1]{\langle #1 \rangle}

\providecommand{\Tr}[1]{\mbox{Tr}\left( #1\right)}

\def \Out{{}}
\def \nfrac{\mathfrak n }

\makeatletter
\def\bbl@set@language#1{%
  \edef\languagename{%
    \ifnum\escapechar=\expandafter`\string#1\@empty
    \else\string#1\@empty\fi}%
  \@ifundefined{babel@language@alias@\languagename}{}{%
    \edef\languagename{\@nameuse{babel@language@alias@\languagename}}%
  }%
  \select@language{\languagename}%
  \expandafter\ifx\csname date\languagename\endcsname\relax\else
    \if@filesw
      \protected@write\@auxout{}{\string\select@language{\languagename}}%
      \bbl@for\bbl@tempa\BabelContentsFiles{%
        \addtocontents{\bbl@tempa}{\xstring\select@language{\languagename}}}%
      \bbl@usehooks{write}{}%
    \fi
  \fi}
\newcommand{\DeclareLanguageAlias}[2]{%
  \global\@namedef{babel@language@alias@#1}{#2}%
}
\makeatother

\DeclareLanguageAlias{en}{english}
\DeclareLanguageAlias{En}{english}
\DeclareLanguageAlias{EN}{english}

\graphicspath{{Images/}}

\begin{document}

% roberts caption package makes all captions centred.
\captionsetup[figure]{justification=raggedright}

%\preprint{}

\title{Characterising particulate random media from near-surface backscattering: a machine learning approach to predict  particle size and concentration} %Title of paper

\author{Artur L Gower}
\email[]{arturgower@gmail.com}
\homepage[]{https://arturgower.github.io/}
%\thanks{}
\affiliation{School of Mathematics, University of Manchester, Oxford Road, Manchester, M13 9PL, UK}

\author{Robert M Gower}
\email[]{gowerrobert@gmail.com}
\homepage[]{https://perso.telecom-paristech.fr/rgower/}
%\thanks{}
\affiliation{LTCI, T\'el\'ecom Paristech, Universit\'e Paris-Saclay, 75013, Paris, France}

\author{Jonathan Deakin}
%\thanks{}
\affiliation{School of Mathematics, University of Manchester, Oxford Road, Manchester, M13 9PL, UK}

\author{William J Parnell}
%\thanks{}
\affiliation{School of Mathematics, University of Manchester, Oxford Road, Manchester, M13 9PL, UK}

\author{I. David Abrahams}
%\thanks{}
\affiliation{Isaac Newton Institute for Mathematical Sciences, 20 Clarkson Road, Cambridge CB3 0EH, UK}


\date{\today}

\begin{abstract}
To what extent can particulate random media be characterised using direct wave backscattering from a single receiver/source? Here, in a two dimensional setting, we show using a machine learning approach that both the particle radius and concentration can be accurately measured when the boundary condition on the particles is of Dirichlet type. Although the methods we introduce could be applied to any particle type. In general backscattering is challenging to interpret for a wide range of particle concentrations, because multiple scattering cannot be ignored, except in the very dilute range. Across the concentration range from 1\% to 20\% we find that the mean backscattered wave field is sufficient to accurately determine the concentration of particles. However, to accurately determine the particle radius, the second moment, or average intensity, of the backscattering is necessary. We are also able to determine what is the ideal frequency range to measure a broad range of particles sizes. To get rigorous results with supervised machine learning requires a large, highly precise, dataset of backscattered waves from an infinite half-space filled with particles. We are able to create this dataset by introducing a numerical approach which accurately approximate{s} the backscattering from an infinite half-space.
\end{abstract}

% \pacs{42.25.Dd}{Wave propagation in random media}
% \pacs{43.20.Fn}{Scattering of acoustic waves}
% \pacs{05.10.Ln}{Monte Carlo methods}

\maketitle %\maketitle must follow title, authors, abstract and \pacs

% \section{Backscattering from random media}
%\label{}

% {\bf What are random media:}
Under close inspection, many materials are composed of small randomly distributed particles or inclusions.
% because the alternative, an organised and regular micro-structure, requires constant upkeep.
% {\bf Applications:}
 So it is no surprise that the need to measure particle properties, such as their average size and concentration, spans many physical disciplines. For quick non-invasive measurements, waves, either mechanical, electromagnetic or quantum, are the preferred choice.
However, measuring a broad range of particle concentrations and sizes is still an open challenge. For high concentration{s} the wave undergoes multiple scattering, which requires specialised methods to compute and interpret. And further, measuring a wide range of particle sizes means a wide range of frequencies needs to considered.

% The size of the particle often dictates the type of wave, because the wavelength needs to be on a similar scale as the particles' dimensions. And the choice between eletromagnetic and mechanical waves also depends on the particles impedance.

The type of wave used depends on the type of particle: acoustic waves are used to measure liquid emulsions~\cite{challis_ultrasound_2005}, sediment on the ocean floor\cite{thorne_overview_2014} and polycrystalline materials~\cite{hu_contribution_2015}. Microwaves are vital in remote sensing of ice~\cite{winebrenner_sea-ice_1989}; optics for aerosols~\cite{torres_derivation_1998} and cellular components, both micrometer~\cite{almasian_oct_2017} and nanoscale~\cite{yi_can_2013} structures, among many other applications.
% \cite{winebrenner_microwave_1992}
% {\bf Backscattering current methods and draw backs.}
% Our focus is on using backscattering from one fixed receiver to measure the medium, such as the device in Figure~\ref{fig:device}.
% However, for highly concentrated particles, it is still an open challenge to create an accurate method of measuring particle properties using waves.
In all these applications, there are cases when transmission experiments are impractical, because
either the material is too opaque or, for example, has an unknown depth. The next natural choice is to use reflected, or \emph{backscattered}, waves.

  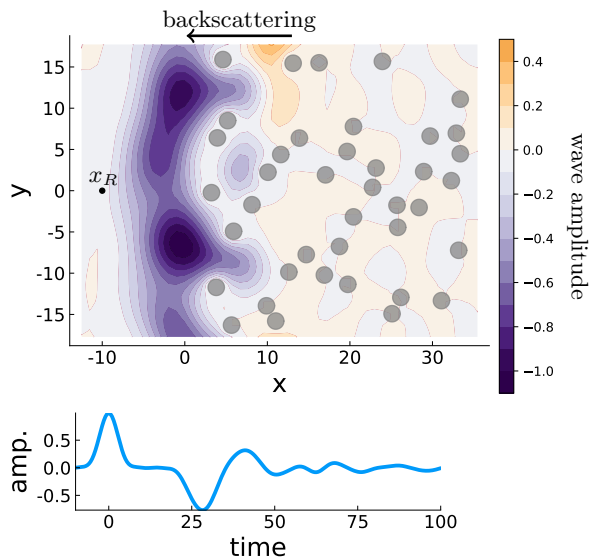
\begin{figure}
    \centering
     \begin{tikzpicture}[line width=1pt]
       \node[inner sep=0pt] at (0,0)
        {
        \includegraphics[width=0.85\linewidth]{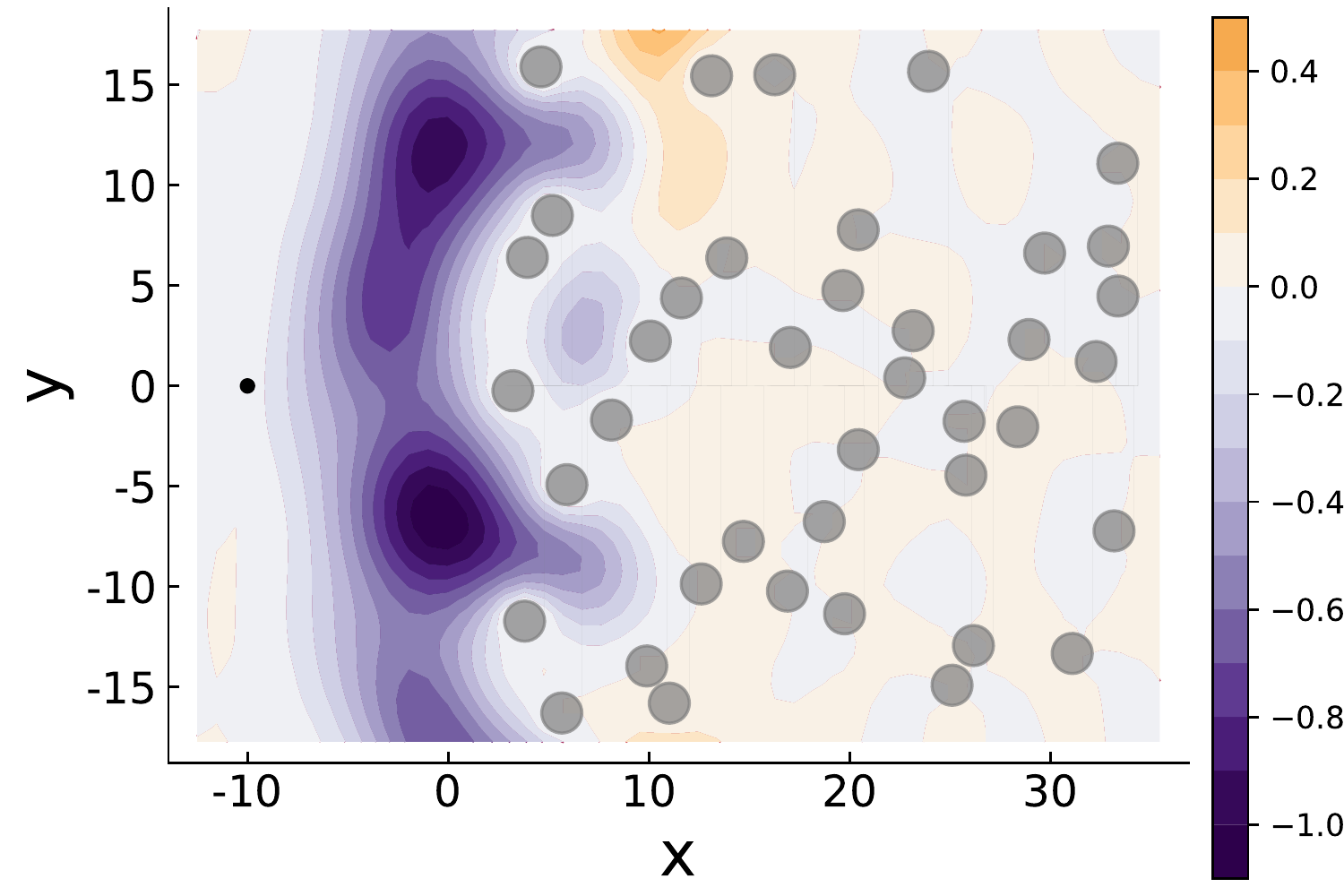}
        };
        \node[rotate=-90] at (4.,0.) {wave amplitude};
       \node at (-2.3,0.5)
        {$x_R$};
       \draw [->] (0.2,2.4) -- (-1.2,2.4);
       \node at (-0.5,2.6)
        {backscattering};
        \node[inner sep=0pt] at (-0.6,-3.6)
         {
         \includegraphics[width=0.7\linewidth]{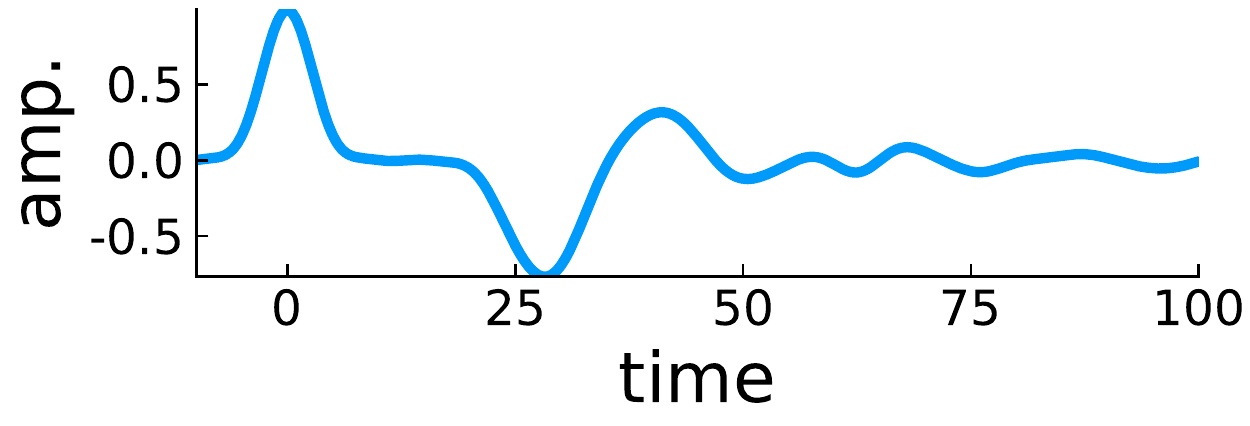}
         };
     \end{tikzpicture}
     \caption{the snapshot above is of a plane wave pulse being backscattered by the grey particles, in the region $x>0$, after time $t=20$ (non-dimensional). The incident pulse originated at the line $x = x_R$, then travelled towards the particles and was then backscattered. The blue line graph shows the amplitude measured at $(x_R,0)$ over time, where around time $t= 25$ the backscattered waves begin to arrive. The particles occupy 10\% of the volume and around 150 particles were used for these simulations.}
     % The particles occupy 10\% of the volume and have radius 1. In practice around 150 particles were used for these parameters. }
     \label{fig:device}
  \end{figure}

Here we ask can one source/receiver measure the properties of a random particulate medium? And is it possible to do so without measuring the backscattering for a range of scattering angles, and without knowing the depth of the medium?
% Though we do assume that it is possible to measure the backscattering for many different particle positions.\rob{$\leftarrow$ I don't understand the use of this ``Though'', and I don't see how the two sentences connect.}
% so that we can not infer the average sound speed or attenuation.

 Figure~\ref{fig:device} illustrates a backscattered wave in time measured at one point in space. We consider only elastic scattering, and scattered waves that have the same frequency as the incident wave.
 % In this case, there is clear evidence that the backscattering intensity from dense media is sensitive to the characteristics of the particles~\cite{tourin_multiple_2000}.
 % so techniques such dynamic light scattering~\cite{pine_diffusing-wave_1990} and .
 % Elastic scattering is often stronger than inelastic sc, and
 We show that, with this simple setup, it is possible to recover a wide range of concentrations and particle radiuses, even including particles with a sub-wavelength radius. We also identify which part of the backscattered signal is sensitive to the concentration and particle radius.
 % Although there is experimental evidence~\cite{weser_particle_2013} showing that direct backscattering is sensitive to particle size and concentration,
 % in the frequency range we use below.
 % Their scatterers are hard, so closer to Neumman boundary and also report their theory relating attenuation to backscattering doesn't really work.
% This is challenging for high concentration, because multiple scattering can not be ignored, and for a wide range of radius’s, because a wide frequency range needs to considered.
To achieve these goals, we use \emph{learning curves} from supervised machine learning, and, in doing so, we also
% Other than investigating the underlying physics, this work has a second goal: to
show how to accurately predict particle radius and concentration from backscattered waves. Supervised machine learning in similar contexts has already shown great promise\cite{rupp_fast_2012,noia_neural_2018}. {See~\cite{noia_neural_2018} for a summary of machine learning applications in remote sensing}.
% Neural Networks and Support Vector Machines and Their Application to Aerosol and Cloud Remote Sensing: A Review

The long term goal is to develop a device, as simple as possible and with little prior information, that can determine the statistical properties of the particles for a broad range of random media.
 To do so will require theoretical predictions, experiments and simulations of backscattered waves.
A supervised learning approach can then easily combine data from these different sources to produce an algorithm that predicts particle statistics. Here we take the first step towards this goal,
by using simulated data, as it is the most accurate for a broad range of media.

The most common approach to determine particle properties from backscattering, to date, is to adjust the parameters of a mathematical model until it fits the measured backscattering~\cite{thorne_overview_2014,weser_particle_2013}. Ideally, these two approaches could be combined to produce an accurate method valid for a large range of parameters.
% Ideally, these approaches could be used to refine the predictions from a supervised machine learning algorithm in a local minimization phase.
% But in terms of the global problem, machine learning can answer what can be measured over a broad range...
%

\emph{\textbf{Simulating near-surface backscattering \textemdash}} We consider a simple setting with Dirichlet boundary conditions, that is, where the scalar wave-field $w =0$ on the boundary of the particles, which for acoustics corresponds to zero pressure, for elasticity corresponds to zero displacement and for electromagnetism corresponds to zero electric or magnetic susceptibility, depending on the polarisation. This case is particularly challenging for many of the current theoretical approaches, as they can lead to unphysical results, even for low frequency and low concentration, as we demonstrate below.
% To extend our results to other types of particles, different boundary conditions would need to be considered.
 We restricted ourselves to two dimensions to lighten the computational load, which is qualitatively similar to three dimensions~\cite{galaz_experimental_2010}. In the conclusion we discuss extensions to three dimensions.
 % and has even been used to quantitatively match 3D experiments~\cite{galaz_experimental_2010}.
% We will determine what moments and frequency range of the signal are necessary.
% {\bf Our simple novel question and method.}

% a snapshot of a backscattered pulse after time $t=20$. Initially at $t=0$ the plane wave pulse was centred at $x = x_R$, then it travelled towards the random medium, the grey particles, and then reflected back to the receiver at $(x,y)=(x_R,0)$. The colour indicates the wave amplitude. The particles occupy 10\% of the volume and have radius 1 (non-dimensional). In practice around 150 particles were used for these parameters.

% \begin{figure}
% \begin{tikzpicture}
%   \node[inner sep=0pt] (0,0)
%     {\includegraphics[height=0.54\linewidth]{../Images/backscatter.pdf}};
%   \node at (-2.2,0.5)
%     {$x_R$};
% \end{tikzpicture}
%  \caption{a snapshot of a backscattered pulse after time $t=20$. Initially at $t=0$ the plane wave pulse was centred at $x = x_R$, then it travelled towards the random medium, the grey particles, and then reflected back to the receiver at $(x,y)=(x_R,0)$. The colour indicates the wave amplitude. The particles occupy 10\% of the volume and have radius 1 (non-dimensional). In practice around 150 particles were used for these parameters. }
%  \label{fig:device}
% \end{figure}

 Consider an incident plane wave $\ee^{\ii k (x -x_R  - t)}$, where $k$ is the wavenumber of the background medium, and we have non-dimensionalised by taking the phase velocity of the background to be 1. We non-dimensionalise because the theory applies to many different applications. % , so as to non-dimensionalise
 The total wave $w = \ee^{\ii k (x -x_R -t)} + u$ satisfies the two dimensional scalar wave equation, where $u$ is the backscattered wave from the particles within the halfspace $x>0$ and $(x_R,0)$ is the receiver position, such as shown in Figure~\ref{fig:device}. If the receiver is close to the particles, then near-field effects will dominate and many realisations will be needed to calculate the statistical moments. To avoid this, we choose $x_R = -10$. We use $a$ for particle radius, $\nfrac$ for number of particles per unit area (concentration) and $\phi = a^2 \pi \nfrac$ for volume fraction, and consider a wide range of media:
 \begin{equation}
  1\% \leq \phi \leq 21\%,  \quad  0.2 \leq a \leq 2.0 \quad \text{and} \quad 0 \leq k \leq 1,
  \label{eqn:parameter_region}
 \end{equation}
 for instance, these values are typically used in emulsions, suspensions, and for atmospheric aerosols.

For random media it is convenient to use the moments of $u$. That is, if $\Lambda$ represents one configuration of particles, then $u = u(\Lambda)$ depends on $\Lambda$ and its ensemble average is $\ensem{u} = \int u(\Lambda) p(\Lambda) d\Lambda$, where $p(\Lambda)$ is the probability of the particles being in the configuration $\Lambda$, then the central moments are
\begin{equation}
\ensem{u}_n = \ensem{ (u - \ensem{u})^n}^{1/n}.
\label{eqn:moments}
\end{equation}
% $\ensem{u}$ is often called the effective backscattering, while $\ensem{u}_2$ is the standard deviation.
% where we consider an ideal receiver, capable of measuring all of these moments.
We will now associate each medium with a fixed particle radius, concentration and set of moments $\ensem{u}$ and~\eqref{eqn:moments}.
 % \ensem{u}_2, \ldots, \ensem{u}_4$.

There are many specialised methods to determine these moments~\cite{martin_multiple_2011,mishchenko_multiple_2006,snieder_coda_2002,garnier_fourth-moment_2016,sheng_introduction_2006}. Those that accurately calculate
 $\ensem{u}_j$ for a broad frequency range require $\ensem{u}$, and a common approximation of $\ensem{u}$ is to assume that $\ensem{u} \approx \ee^{\ii k_* x}$ inside the random media, for some effective wave number $k_*$. For small volume fraction $\phi$ and direct backscattering~\cite{foldy_multiple_1945,gower_reflection_2017} this approximation leads to
$
\ensem{u} \approx  - \ii \phi (  \pi a^2 k^2)^{-1} \sum_n J_n(k a)/H_n(k a) \ee^{\ii (n \pi -  k x)},
$
where $J_n$ and $H_n$ are a Bessel and Hankel function of the first kind.
 % $\ensem{u} \approx  \ii \phi (  \pi a^2 k^2)^{-1} f(\theta) \ee^{- \ii k x}$, where $f(\theta) = - \sum_n \ee^{\ii n \theta} J_n(k a)/H_n(k a)$.
However, this approximation diverges when $a \to 0$, while $\phi$ is fixed, and leads to the unphysical result $|\ensem{u}|>1$.
Even rigorous methods\cite{martin_multiple_2011}, deduced for moderate volume fraction, present the same problem. This problem is a result of strong scatterers, with $w = 0$ on their boundaries, completely reflecting waves at low-frequencies for any particle volume fraction. This can be seen by investigating the effective properties~\cite{parnell_multiple_2010,martin_estimating_2010}. The consequence is that series expansions of $\ensem{u}$ for small volume fractions do not converge for scatterers with $w = 0$ on their boundaries. For other strong scatterers with $w \approx 0$ on their boundaries, this series converges very slowly.
% This can be seen by examining the effective bulk modulus $\beta_\eff$ of a composite~\cite{parnell_multiple_2010}: $1/\beta_\eff = (1-\phi)/\beta + \phi/\beta_o$, where $\beta$ and $\beta$ are respectively

% \begin{figure}
% \centering
%  \includegraphics[width=0.6\linewidth]{EffectiveReflectionAmplitudeVol-k0p1.pdf}
%  % \includegraphics[width=0.64\linewidth]{EffectiveReflectionAmplitudeConc-reg-k0p3.pdf}
% % \flushleft \vspace{-0.7cm}
%  % \hspace{0.0\linewidth} $a)$ \hspace{0.44\linewidth} $b)$
%  \caption{the density plot of $|\ensem{u}|$, the absolute value of the mean backscattering, for $k = 0.1$ and incident wave $\ee^{\ii k(x- x_R - t)}$ predicted by~\cite{martin_multiple_2011}. The white regions are where $|\ensem{u}| > 1$, which is unphysical.}
%  \label{fig:mean_analytic}
% \end{figure}

% So in light scattering, often \ensem{u_R}_2 is easier to measure. The oscillations are so fast that \ensem{u_R} gets averaged to zero over time. Alternatively, it is possible to measure \ensem{u_R} and \ensem{u_R}_2 in some acoustics and other electromagnetic settings

%  we perform many simulations for the same radius $a$ and volume fraction. If each of these time responses is a vector $\vec X_i$, then the moments $\vec M_i$ become
% \begin{equation*}
% \vec M_1 = \frac{1}{P}\sum_{i=1}^N \vec X_i, \quad \vec M_j = \left[ \frac{1}{P-1}\sum_{i=1}^P (\vec X_i - \vec M_1)^j \right ]^{1/j},
% \end{equation*}
% where the powers $j$ and $1/j$ are taken element wise, and the $\vec M_j$ have the same length scale.
\begin{figure}
\centering
 \includegraphics[width=0.49\linewidth]{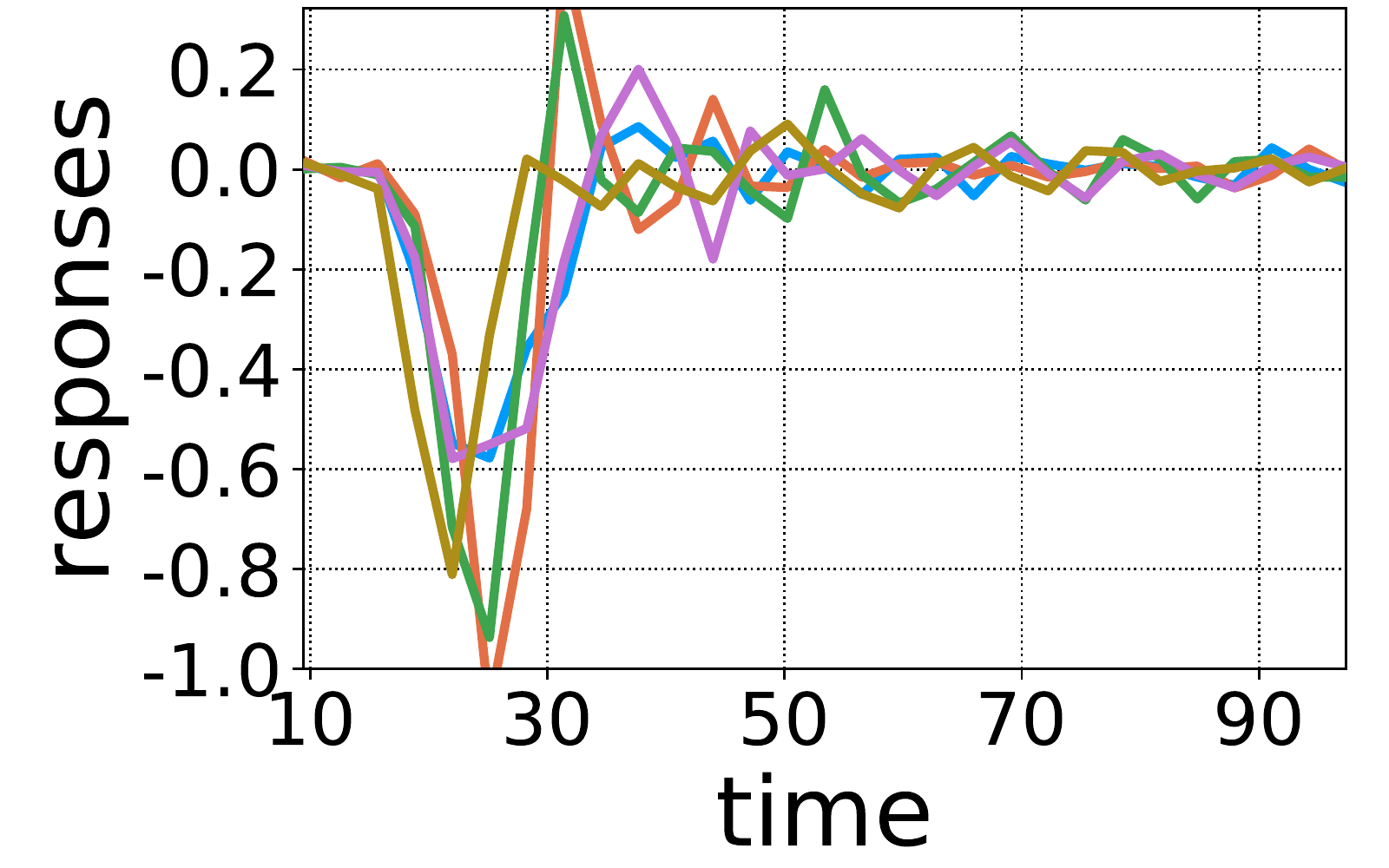}
 \includegraphics[width=0.49\linewidth]{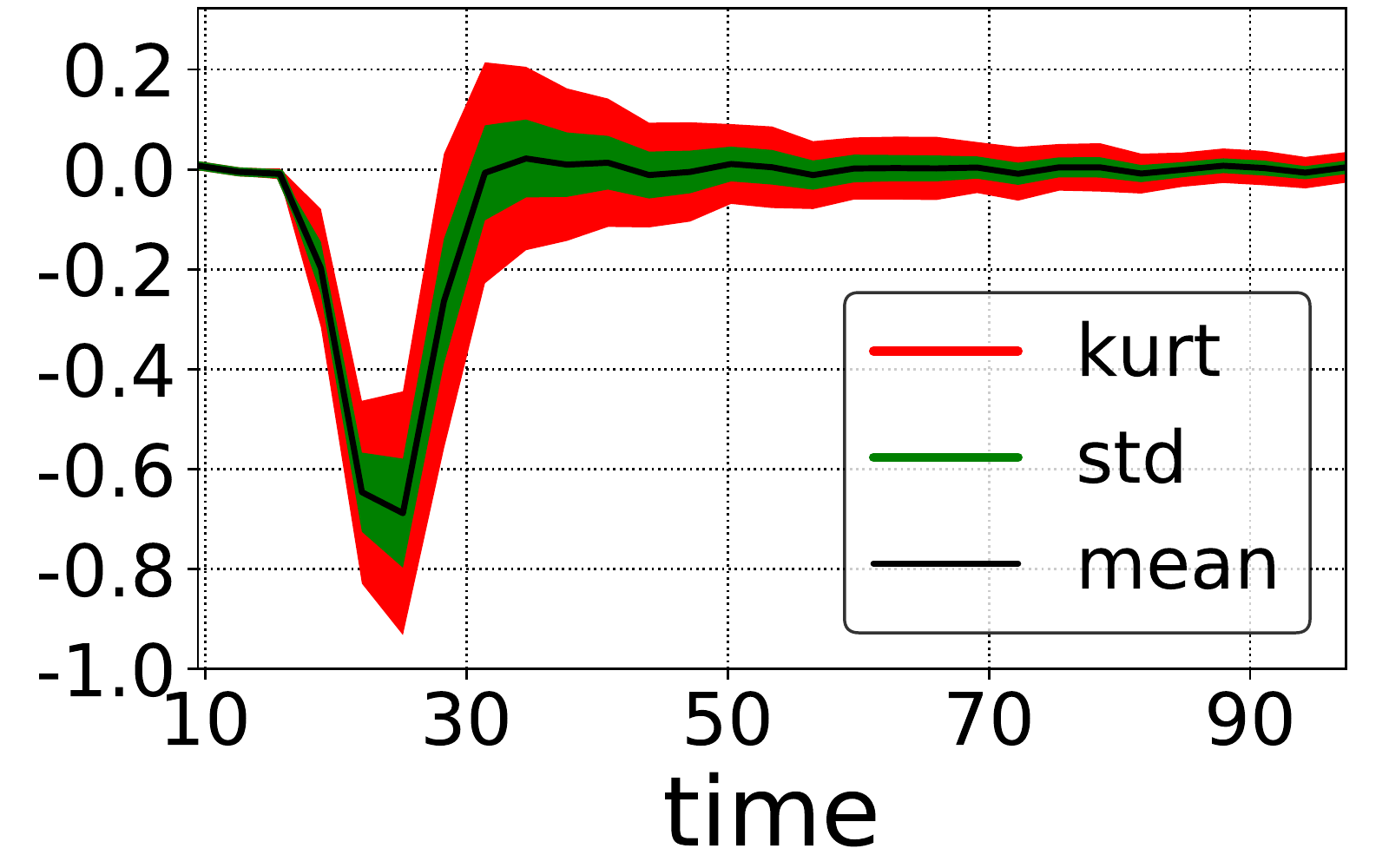}
% \flushleft \vspace{-0.7cm}
 % \hspace{0.0\linewidth} $a)$ \hspace{0.44\linewidth} $b)$
 \caption{the backscattering of the incident wave $\ee^{-0.1(x - x_R -t)^2}$ from particles of radius $a= 0.2$ occupying $ \phi = 20\%$ of the volume.  Left, backscattering from five different configurations. Right, the moments of 756 configurations. The height of the black line is $\ensem{u}$ the mean response, while the total thickness of the green and red regions are the second $\ensem{u}_2$ (standard deviation) and fourth $\ensem{u}_4$ (kurtosis) moment.}
 \label{fig:responses}
\end{figure}

It may be possible to accurately describe backscattering from strong scatterers with integral methods that are valid for any volume fraction~\cite{cobus_dynamic_2017,kristensson_coherent_2015}. Though, we note, that methods dervived from Lippmann-Schwinger type equations are not formally valid for scatterers with discontinuous material properties~\cite{martin_acoustic_2003}, such as strong acoustic scatterers.

% \subsection{Simulating backscattering}
To accurately determine all the moments over the range~\eqref{eqn:parameter_region}, we use a numerical approach based on the multipole method~\cite{martin_multiple_2006} to calculate $u(\Lambda)$ for each configuration $\Lambda$, from which we determine the $\ensem{u}_n$ {with a Monte Carlo method}\footnote{The alternative would be to piece together different theoretical methods, whose range of validity is not clear~\cite{tourin_multiple_2000}.}. In all our convergence tests, truncation errors and benchmarks were within 1\% accuracy for each simulation. In the supplementary material we explain how to reproduce our results, including high performance software \href{https://github.com/jondea/MultipleScattering.jl/tree/v0.1}{to simulate the backscattering}~\cite{art_gower_2017_1133989} and \href{https://github.com/gowerrobert/MultipleScatteringLearnMoments.jl}{implement the machine learning}. {The data used in this paper is also publicly available~\cite{gower_dirichlet_data_2018}.}

% {\bf Near surface and  moments.}
Approximating the backscattering from an infinite halfspace, with a limited computational domain, is challenging.
To overcome this challenge we calculate the backscattering of the incident time pulse $\ee^{-0.1(x - x_R -t)^2}$, which, for wavenumbers $0\leq k \leq 1$, results in less than 1\% Gibbs phenomena, and receive the backscattering at $(x_R,0)$. By only receiving the signal for
% and retain only the part of the signal which was produced by the computational domain. In more detail, we calculate the backscattering
$ t \leq 98$, we can exclude from the simulation all particles that would take more than $t=100$ for their first scattered wave to arrive at the receiver $(x_R,0)$. That is, we need only simulate particles that are near the surface, which is why we call this \emph{near-surface backscattering}.
% \rob{Strange excluding those whose signals return after $t=100,$ yet calculating upto $t=98$. Are we ok with signals that return at $t = 99$?} As a result, the computational domain is only the near-surface region.
% And as $x_R = -10$, the earliest backscattered wave arrives at $t =16$ and is from particles with radius $2$.
 % So we use a discrete Fourier transform to calculate the backscattering for $9.5 \leq t \leq 98$, with a time step $dt \approx \pi$, which leaves us with 29 points on each backscattered wave.
 % , at which point $u(x_R,0, t) = u_b(x_R,0,t)$ because the incident wave has already propagated away from $x = x_R$.
 % And as the largest wavenumber is $k=1$, the smallest time step for the discrete Fourier transform is $dt \approx \pi$, which leaves us with 29 points on each backscattered wave.
 % which we will show is more than enough.
 See Figure~\ref{fig:responses} for the incident time pulse and for $\ensem{u}$, $\ensem{u}_1$ and $\ensem{u}_4$, where we include $\ensem{u}_4$ as it is known to be sensitive the micro-structure\cite{ammari_mathematical_2013,sheng_introduction_2006}.

\begin{figure}
\centering
 \includegraphics[width=\linewidth]{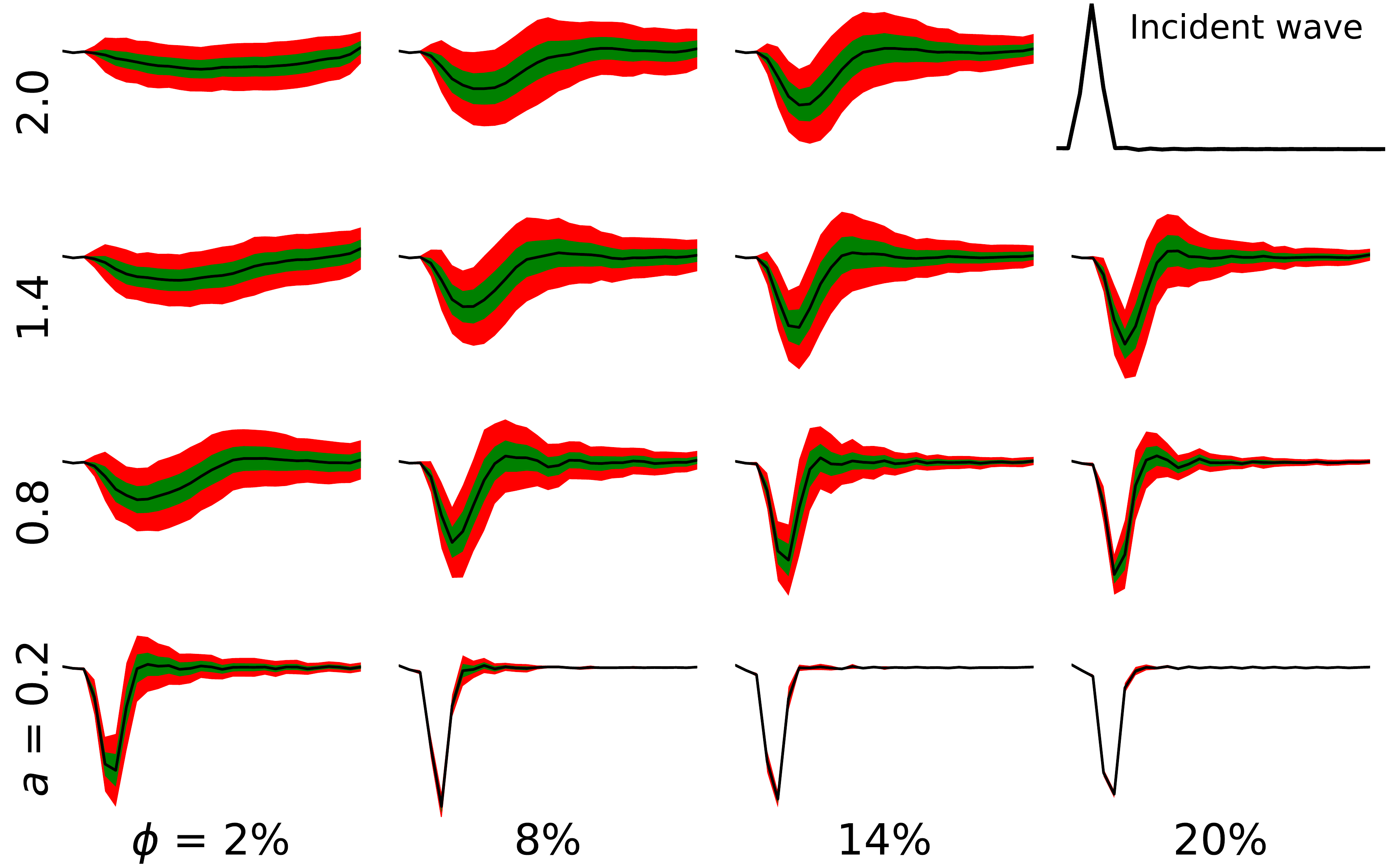}
 \caption{
 an overview of the moments of the direct backscattering of the incident wave shown in the top-right. Each graph shows $-1<y<0.35$, while the $x-$axes shows time $9.5 \leq t \leq 98$. Each column has the same volume fraction $\phi$, while each row has the same particle radius $a$, except in the top-right which is shown to the same scale as the moments, but with time $-9.5\leq t \leq 78$ and $-0.35<y<1.0$.  }
 \label{fig:allmoments}
\end{figure}

 In total we simulated the moments of 205 different media, evenly sampled from~\eqref{eqn:parameter_region}, which required 83000 backscattering simulations - each corresponding to one configuration $\Lambda$. For the larger simulations up to 7600 particles were used. To estimate the quality of the calculated moments, we used the standard error of the mean. See Figure~\ref{fig:allmoments} for an overview of the simulated moments.

% \subsection*{Learning from backscattering } \label{sec:learn}

% \art{This is just me thinking aloud. Outline of section:\\
% $\circ$ Why machine learning? And how our methods can predict any polynomial relationship and have been kicking ass since 1990. A blah about training them. I don't think we should start with our conclusions, because we already state them in the abstract and can, if we want, repeat them in the into. This letter is so short, I think it best if it was a continuous story that at every step convinces the reader more and more that are results are solid, rather than just restating our results.
% \\
% $\circ$ present best results, the method works! \\
% $\circ$ But what moments are necessary?\\
% $\circ$ and how can we guarantee that poor quality data isn't affecting our results? Bam, learning curves.\\
% $\circ$ Determine minimum frequency range to get our awesome results.
% $\circ$ Finally, many physicist will want to know: what part of the signal is important? This is where moments of moments could come in, if we had a clear message to give.
% }
% \pagebreak

\emph{\textbf{Learning from backscattering \textemdash}} With a high quality data set of backscattered waves, we can now use supervised machine learning to generate a \emph{model} that best fits the radius and a separate model that best fits the concentration. To test these models, we use them to predict the concentration and the radius of yet unseen media using only backscattered waves as input.
 Our supervised machine learning method of choice is kernel ridge regression~\cite{Boser:1992,Aizerman67theoretical},
 % since the early 90's~\cite{Boser:1992}\footnote{Though first introduced to the machine learning community in the late 60's~\cite{Aizerman67theoretical}}.
 % which is very effective for fitting continuous processes,
because when using continuous kernels it
 can fit any continuous function~\cite{MicchelliXZ06}.
% The effectiveness of kernel ridge regression stems from its expressive power. Indeed, ridge regression models based on continuous kernels
% can fit any continuous function~\cite{MicchelliXZ06}.
This allows us to establish whether the radius or the concentration are continuous functions of $\ensem{u}$ or $\ensem{u}_2$ or both. In other words, we can
determine which moments are needed to predict the radius and concentration. We present the results for concentration, instead of volume fraction, because it can be accurately predicted from just $\ensem{u}$.
% This makes kernel regression the perfect tool to investigate whether the radius or the concentration are functions of given moments of backscattered wave. That is, this will allow us to establish which moments are needed to predict the radius and concentration.

Our \emph{training} set is the simulated backscattered moments of 205 different media. Using this training set we \emph{train} a model, that is to say, we use kernel ridge regression applied to the training set to generate a model.  The hyperparameters of the ridge regression were selected using a 7-fold cross validation.
 To determine the \emph{predictive power} of our model, we generate a \emph{test set} with 81 randomly chosen media with radius $0.2 \leq a \leq 2.0$ and volume fraction $1\% \leq \phi \leq 21\%$. Every medium of the test set is distinct from the training set.
To measure the goodness of fit, we use the $R^2$ coefficient with respect to the mean of the test set. If $R^2 =0$ then the model has the same predictive power as the mean of the test set, while $R^2 =1$ shows that the model has perfect prediction. Finally we tested two continuous kernels, the Gaussian (or radial basis) and the Ornstein--Uhlenbeck kernel. Both kernels gave similar scores
through crossvalidation, though the Ornstein--Uhlenbeck kernel had a slightly better $R^2$ coefficient on the test set, so we only report these results.
  % Thus, in hinde sight, the Ornstein--Uhlenbeck kernel resulted in a better model, so we only report these results.

\begin{figure}
	\centering
	\includegraphics[width=0.49\linewidth]{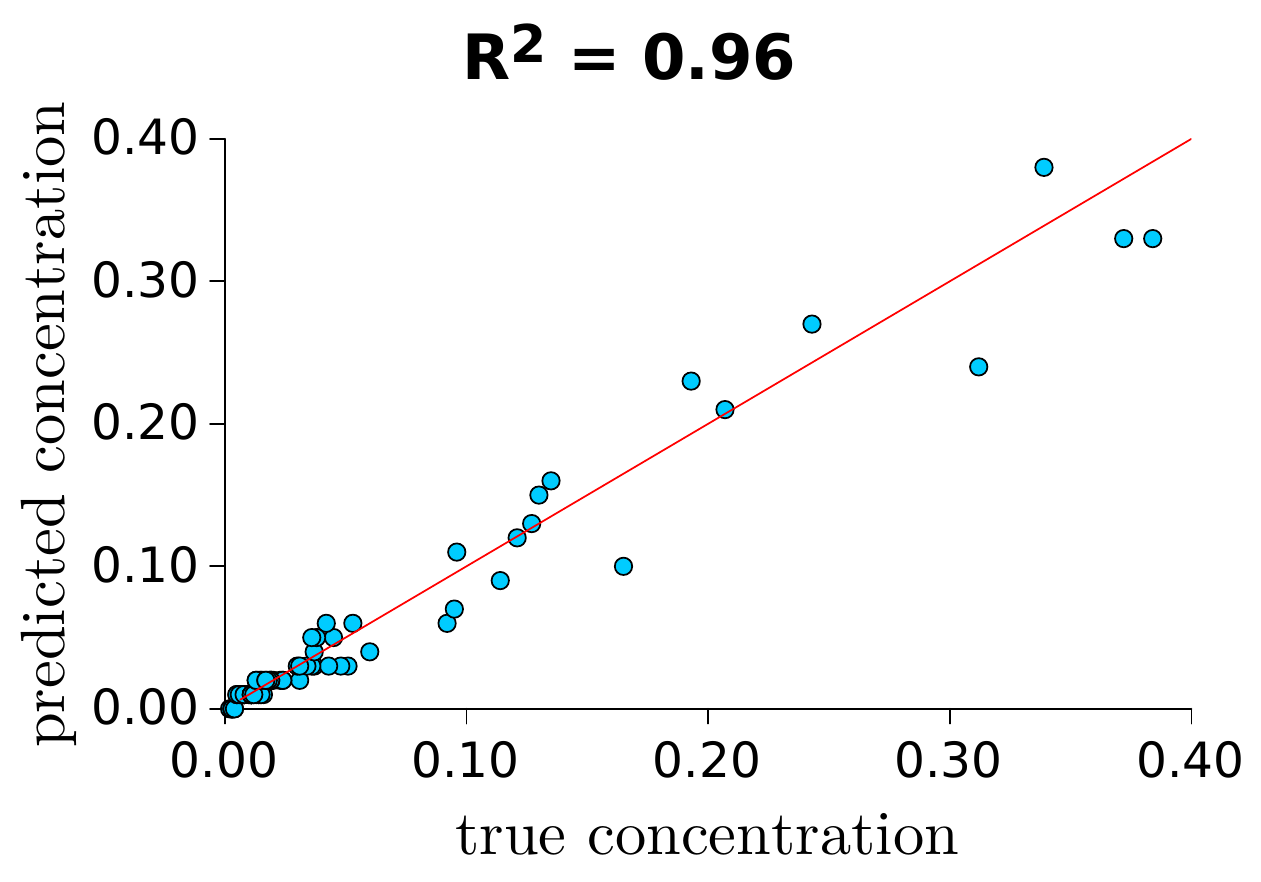}
	\includegraphics[width=0.49\linewidth]{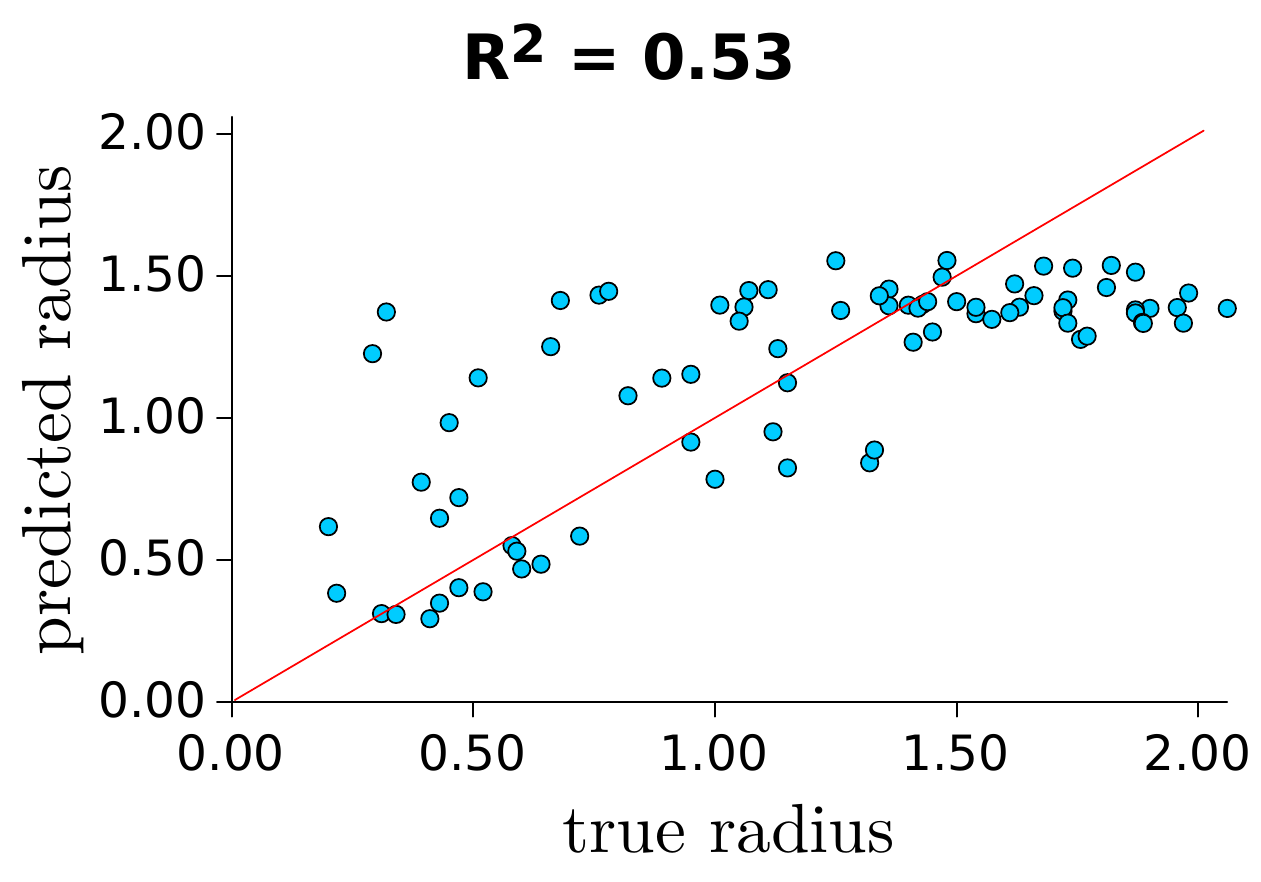}
	% \label{fig:Rsqrmean}
% \end{figure}
% \begin{figure}
	\centering
	\includegraphics[width=0.49\linewidth]{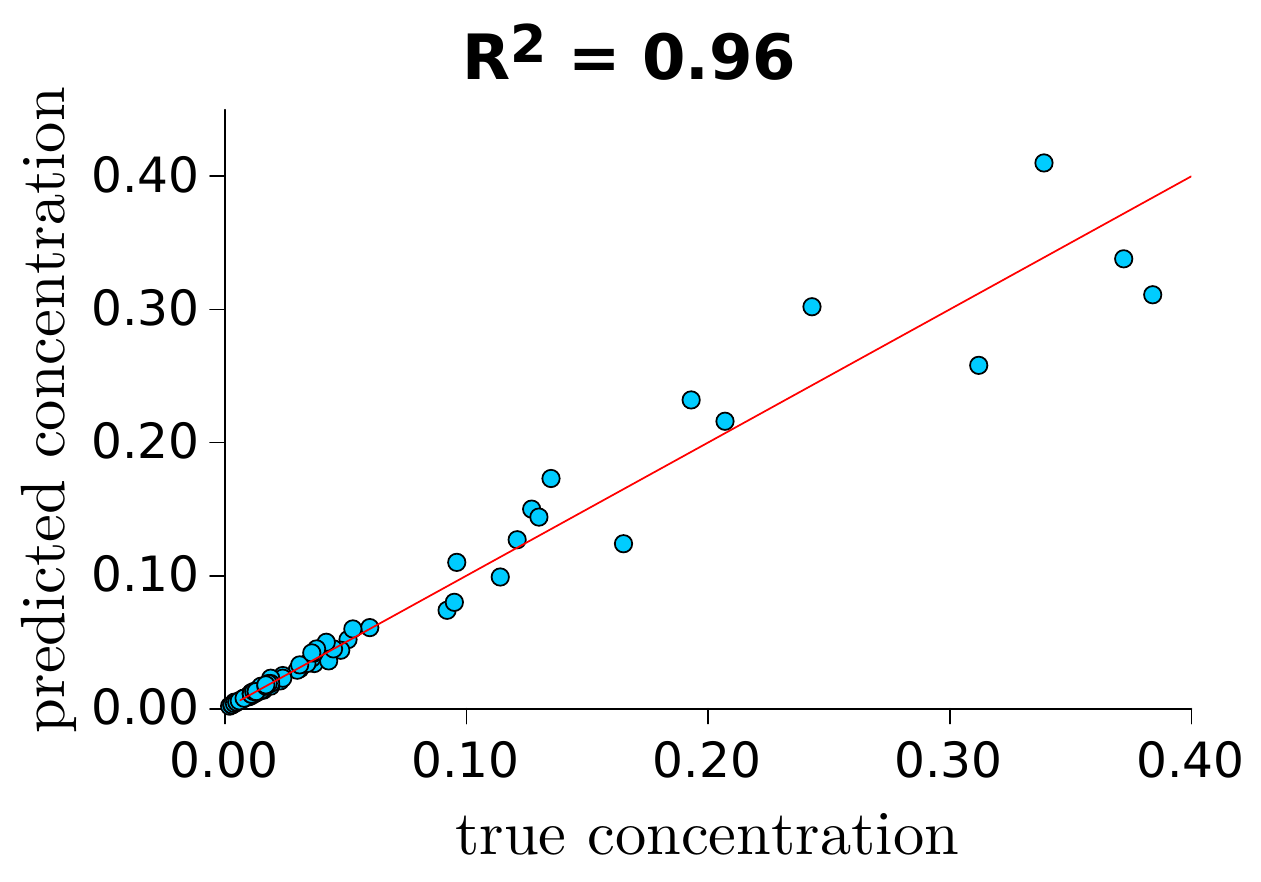}
	\includegraphics[width=0.49\linewidth]{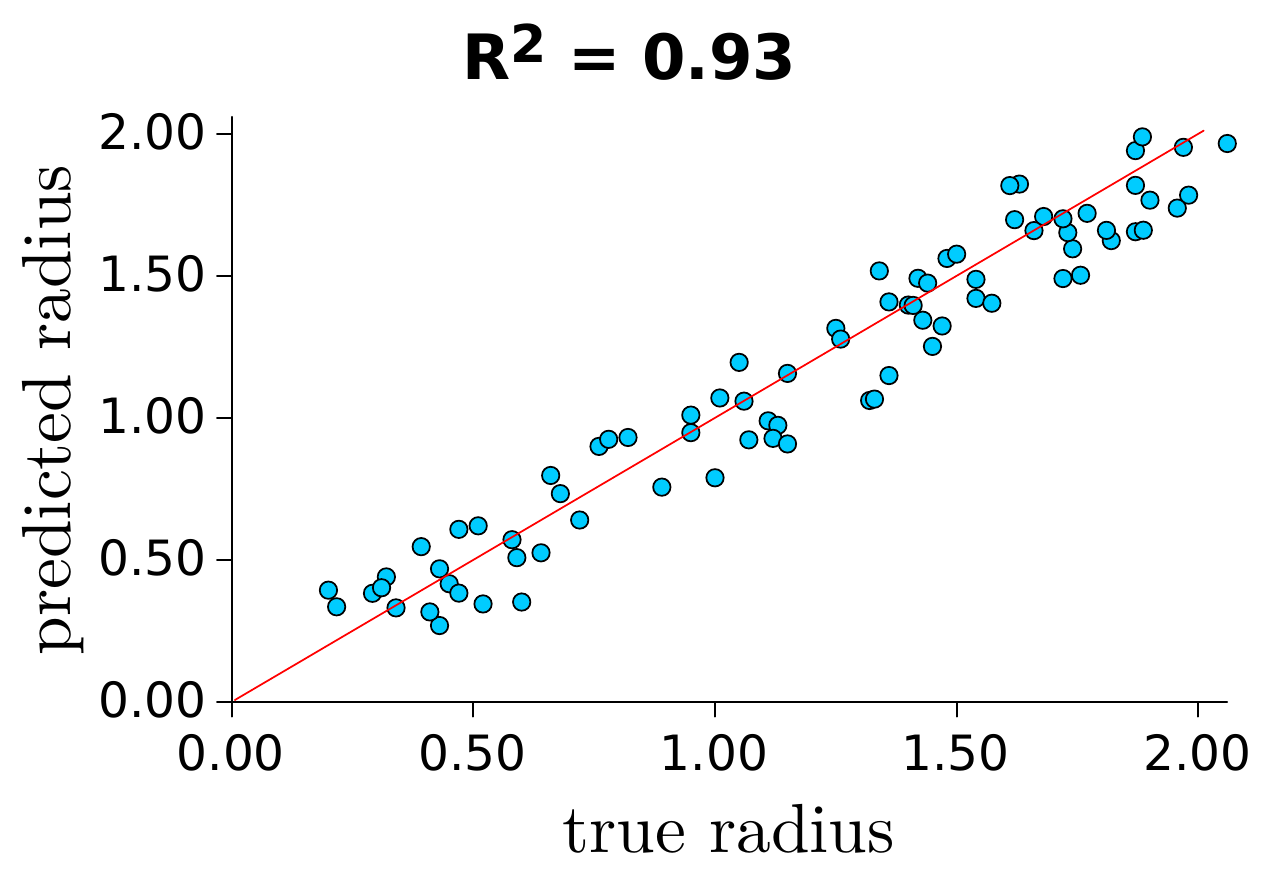}
	\caption{ shows that to accurately predict concentration requires only $\ensem{u}$, but to accurately predict the particle radius requires also second moment $\ensem{u}_2$. The top two models were trained using only the mean $\ensem{u}$, while the bottom two were trained using the mean $\ensem{u}$ and second moment $\ensem{u}_2$. The best prediction for the concentration gives $R^2 =0.98$, which results from using low wavenumbers, discussed later.
		% \art{might add to inset pictures showing what the true and predicted media look like, with arrows to the corresponding labels. Good idea?}
    % \art{Actually, inset would be too small. Would have to be a seperate figure comparing pictures of particles predicted x true, as Rob's original suggestion. }
    }
	\label{fig:Rsqrmom}
\end{figure}

\emph{\textbf{Results \textemdash}} we train two models using only $\ensem{u}$,
one to predict the concentration and one to predict the radius, see the top graphs of Figure~\ref{fig:Rsqrmom}. The top left and top right graphs show the scatter plot of the concentration and radius of the test set against the predicted concentration and radius, respectively.
 The prediction for the concentration is almost perfect, with $R^2=0.96.$ On the other hand, the prediction for the radius is almost meaningless with $R^2 = 0.53.$ The failure of the first moment $\ensem{u}$ alone to predict the radius is significant, as it indicates that the radius is not a continuous function of $\ensem{u}$.
% , indeed, it may simply not be a function of the first moment alone.

To accurately predict the radius,  the second moment was necessary. Indeed, training a model on the first and second moment resulted in an accurate prediction of the radius with $R^2 = 0.93$, see the bottom right of Figure~\ref{fig:Rsqrmom}.

%
%In the supplementary material we complement these experiments by testing the stability of the models and plotting the learning saturation curves.

% To better understand if our conclusions extend beyond our given data set, we investigate the stability of our models to changes in the data.
To show that our results, such as the top right of Figure~\ref{fig:Rsqrmom},
 are not due to insufficient data, and likely extend beyond our data set, we examine the learning curves.
% To validate our models, and to better understand if our conclusions extend beyond our given data set, we will plot their \emph{learning curves}.
A \emph{learning curve} shows the $R^2$ coefficient as the quality of the data is increased.
%by gradually increasing the data quality and recording the progression of the $R^2$ coefficient.
For example, if it was possible to predict the radius from only $\ensem{u}$,
% a model is able to completely predict a property, such as the radius,
then the model's $R^2$ coefficient would increase when improving the training data's quality.
 % until reaching $R^2 =1$.
 Contrary to this, if the $R^2$ coefficient does not increase, or if there is no clear trend, then the model cannot predict the radius, no matter the quality of the training data.
 % And as we are using kernel models, capable of learning any continuous function, these learning curves can reveal which moments can be used to predict the media's properties.
% If a model is successfully learning the underlying phenomena,  we should see a gradual improvement in its predictive power with the $R^2$ coefficient converging to $1$ as we improve the quality or increase the amount of training data. Contrary to this, if the increase in the $R^2$ coefficient saturates before reaching $R^2 =1$ or worst still, there is no clear trend, we know that the model is incapable of learning the underlying process.

% No longer relevant -->
%In particular, a learning algorithm does not overfit if and only if it is stable to small changes in the training data, see Theorem 13.2 in~\cite{Shalev-Shwartz:2014book}.

\begin{figure}
	\centering
	\includegraphics[width=0.49\linewidth]{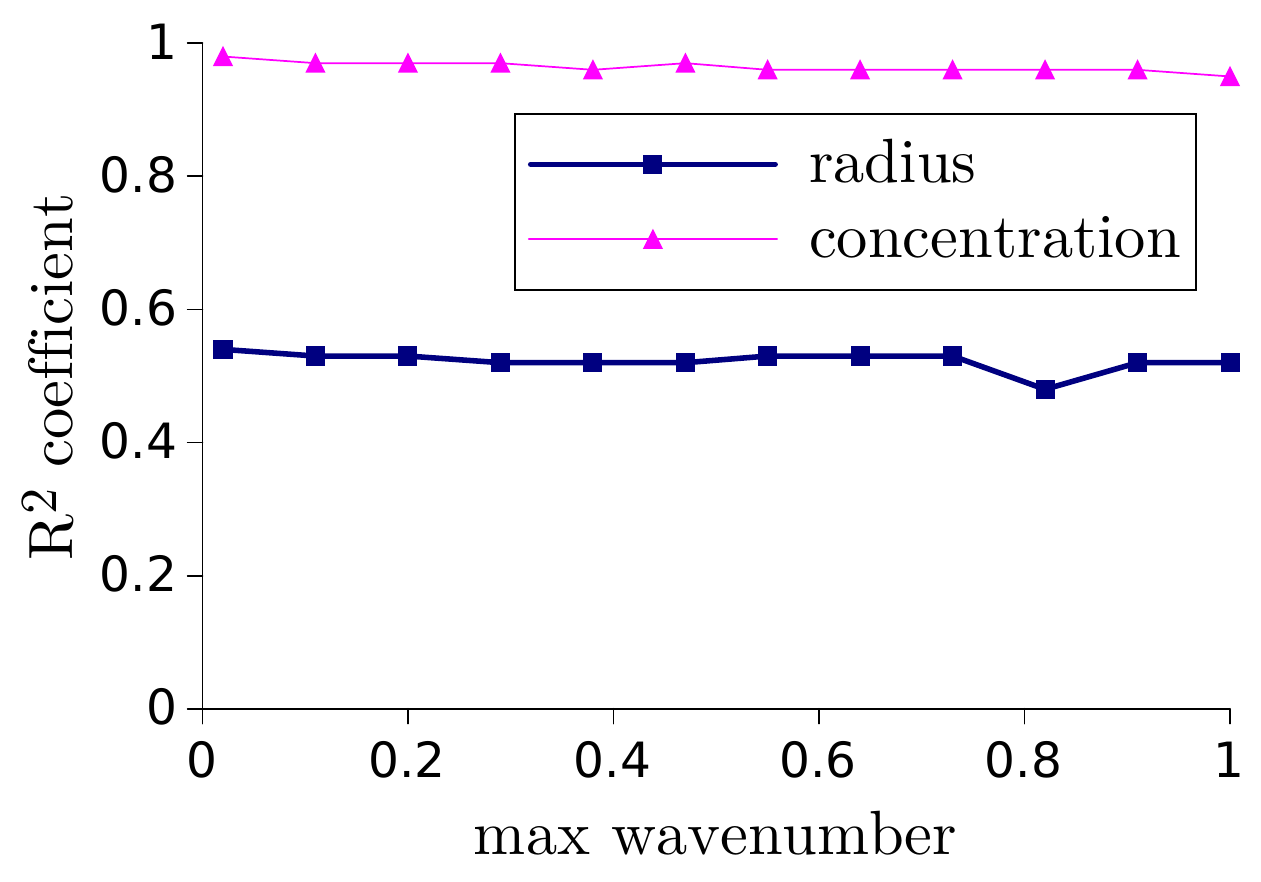}
	\includegraphics[width=0.49\linewidth]{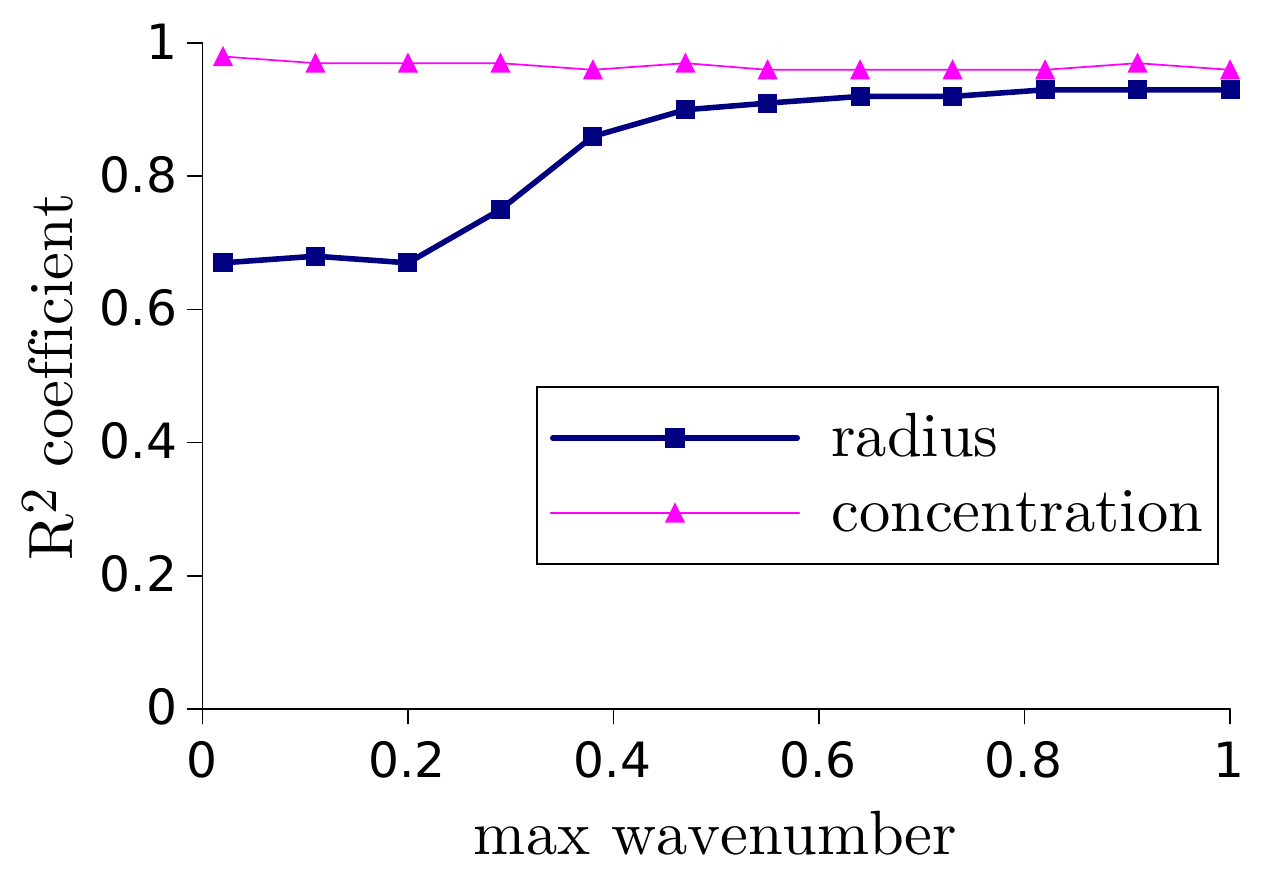}
	\caption{
	shows how increasing the maximum wavenumber does not lead to better predictions of particle radius when measuring $\ensem{u}$. That is, for each point $(x,y)$ on the graphs, we limit the incident wavenumbers of the training and test set to $0 \leq k \leq x$, which results in $y = R^2$.  On the left (right) we used a model trained on only $\ensem{u}$ ($\ensem{u}$ and $\ensem{u}_2$).
% The $R^2$ coefficient of the radius and concentration as we increase the maximum wavenumber $k$ in the training and test set.
}
	\label{fig:stabfreq}
\end{figure}

\begin{figure}
	\centering
	\includegraphics[width=0.49\linewidth]{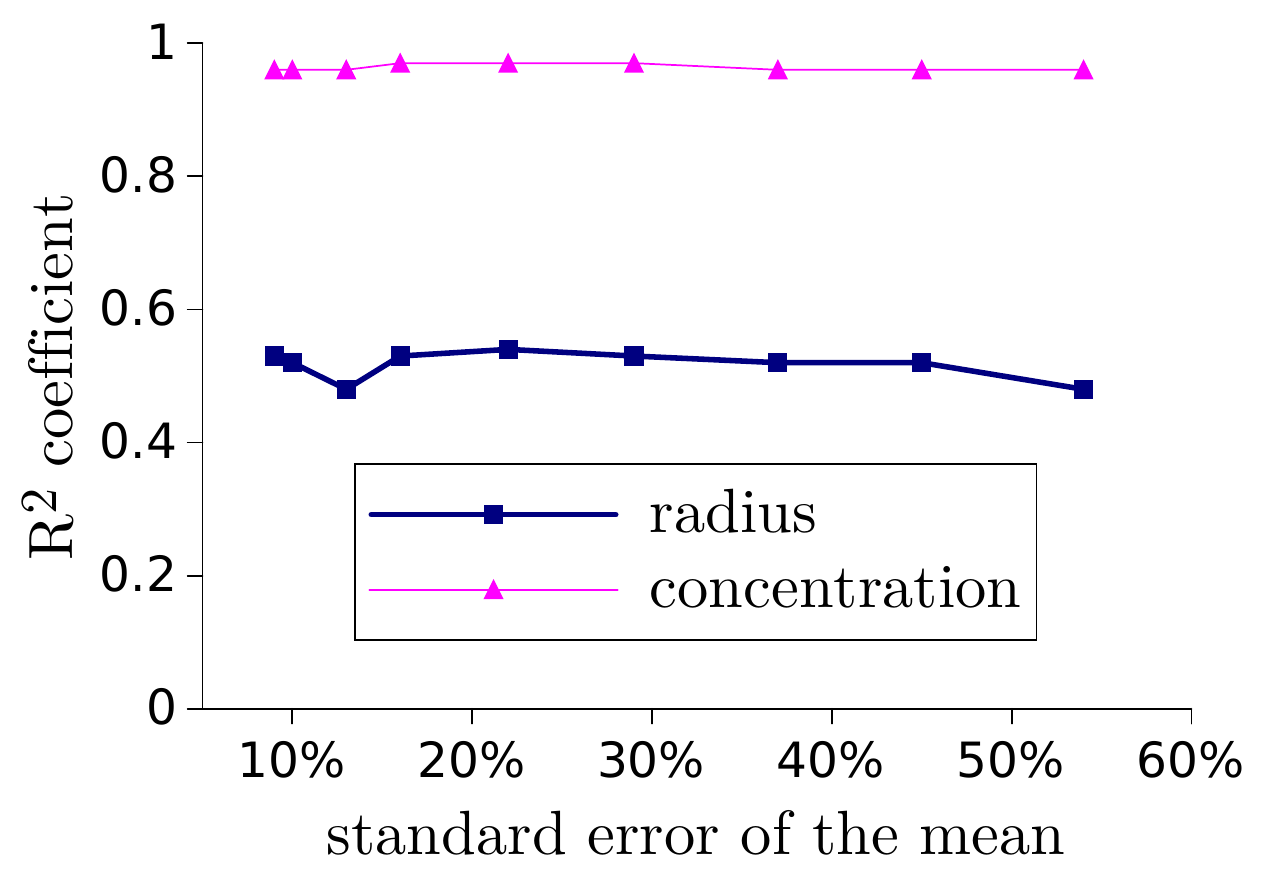}
	\includegraphics[width=0.49\linewidth]{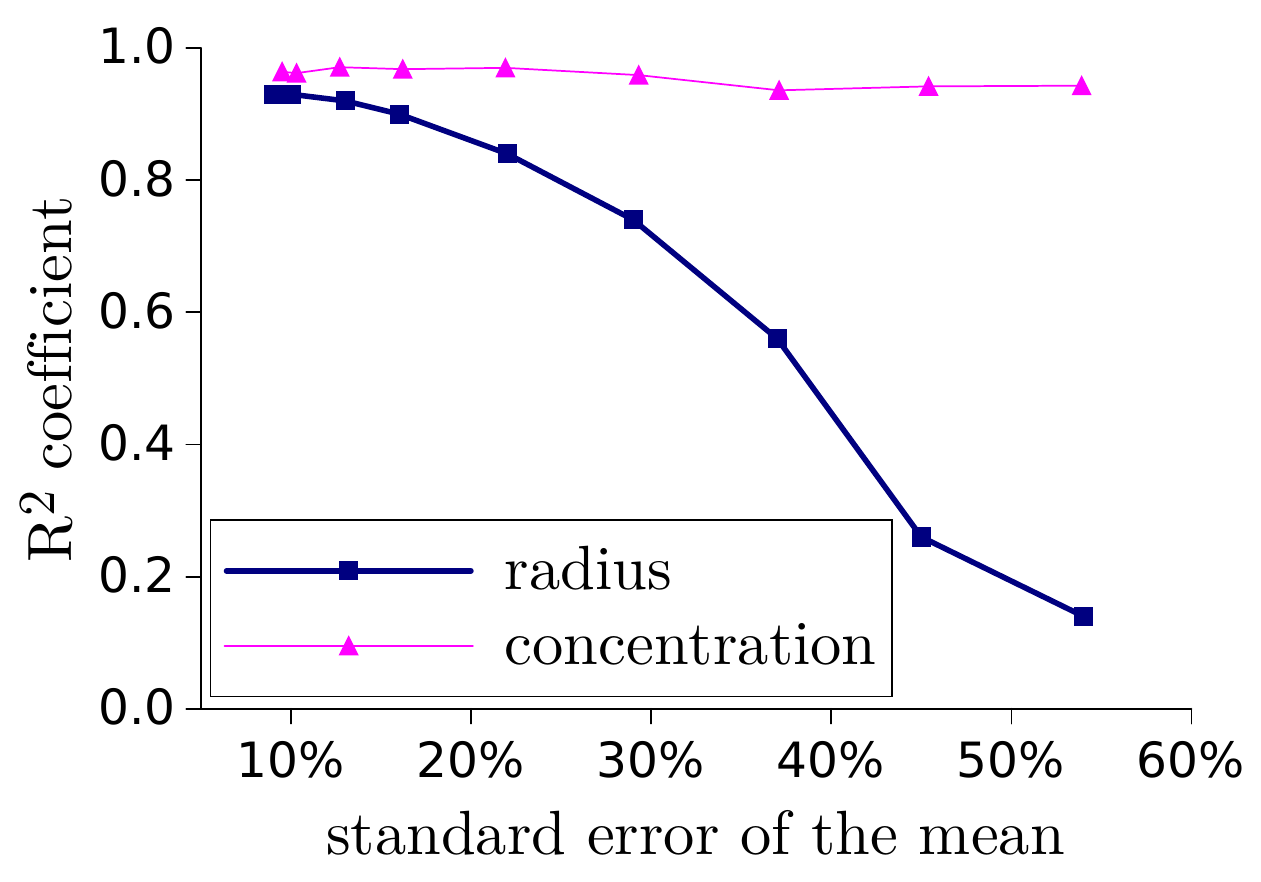}
	\includegraphics[width=0.49\linewidth]{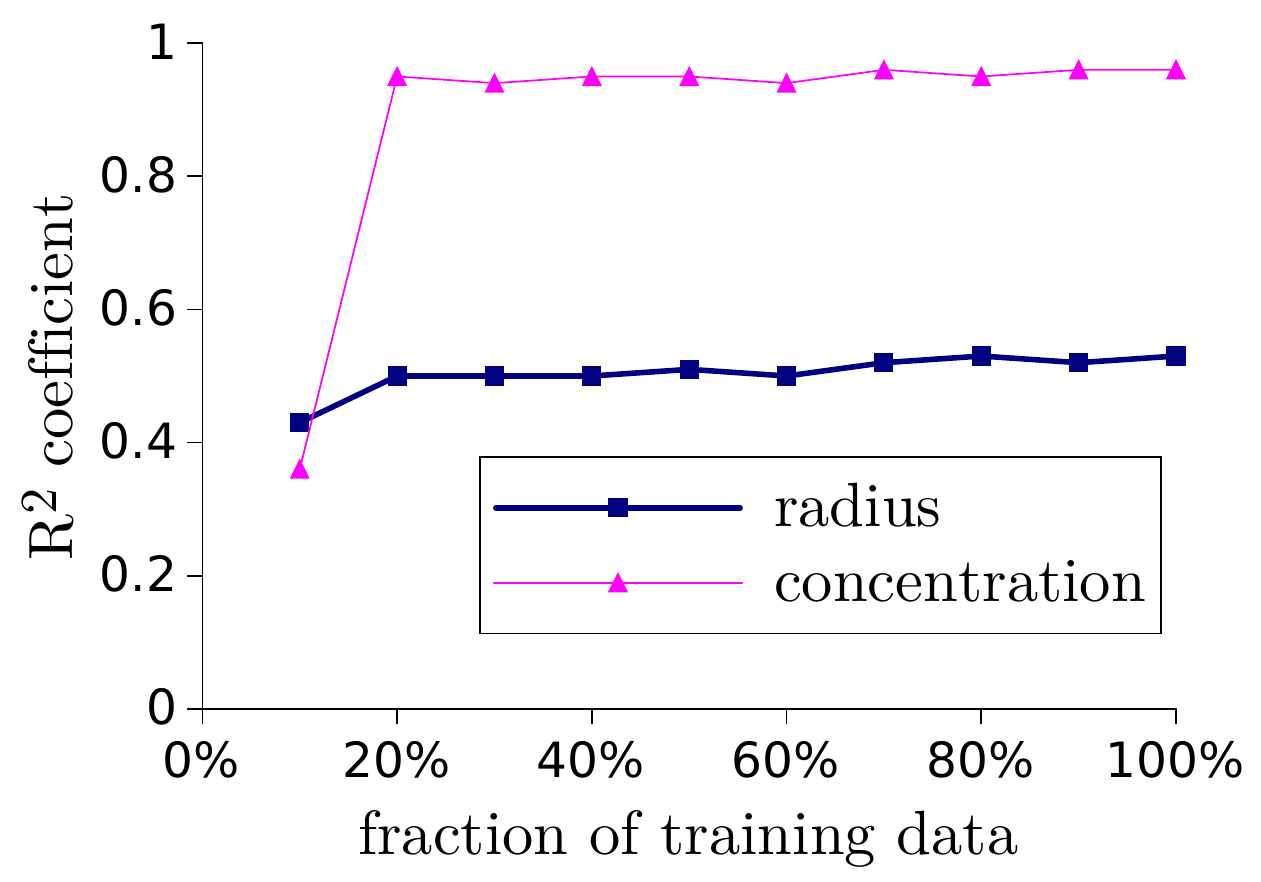}
	\includegraphics[width=0.49\linewidth]{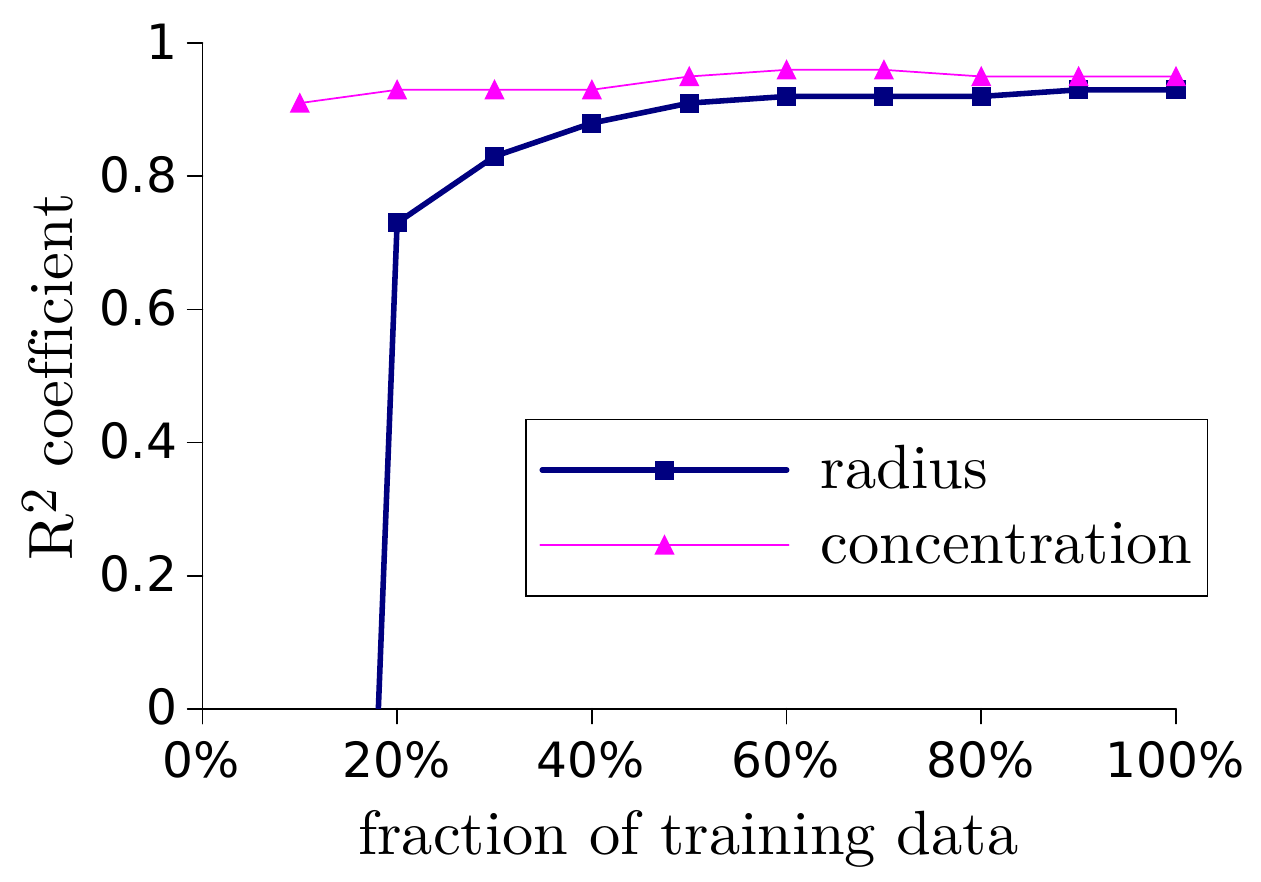}
	\caption{
shows how well the particle radius and concentration are predicted, $R^2$, when changing
the quality of the training data. The test set was fixed with a relative standard error of the mean of $10\%$. The top two graphs increase the number of simulations per medium, resulting in a change of the relative standard error of mean $\ensem{u}$ on the $x$-axis.
The bottom graphs increase the number of media, shown as a percentage of the full training data on the $x$-axis. The model on the left (right) was trained using only $\ensem{u}$ ($\ensem{u}$ and $\ensem{u}_2$).}
	\label{fig:stabnummed}
\end{figure}

%\art{This bit is good, but needs dumbing down. It needs to more clearly say: these learning curves estimate what would happen with perfect data. And indicatee that are results are not a result of poor data because.. blah about the stability and theorem to back it up. }
We vary the quality of the training data by changing: the number of media, the number of simulations for each medium, and by limiting the maximum wavenumber $k$ of the incident wave. For every change in the training data we re-train the model  of the radius and the model of the concentration.
The resulting learning curves are shown in Figures~\ref{fig:stabfreq} and \ref{fig:stabnummed}.
 % For Figure~\ref{fig:stabfreq} alone, we also limited maximum wavenumber of the test data as well as the training data.
% In more detail, $x = k_i$ in Figure~\ref{fig:stabfreq} corresponds to using a training set where the incident wavenumbers are the range $0<k< k_i$.
The graphs on the left of all these figures are the result of using a model trained only on $\ensem{u}$, and from them we see that the $R^2$ of the radius model does not tend to 1 when increasing the training data quality.
The simplest explanation for this is that $\ensem{u}$ does not by itself carry information about the radius.
% This shows that niether increasing the number of media in the training set, nor increasing the number of simulated curves, nor changing the frequency of the response waves will improve the prediction. The simplest explanation for this is that the information content is not there, that is, the first moment does not contain information on the radius.
On the other hand, the graphs on the right of Figures~\ref{fig:stabfreq} and \ref{fig:stabnummed} are models trained on $\ensem{u}$ and $\ensem{u}_2$, and clearly their $R^2$ for the radius converges to $R^2 =1$.
In contrast, the concentration is accurately predicted from $\ensem{u}$ even when using either 30\% of the number of training media, having large standard errors of the mean, or using only wavenumbers $k\leq 0.1$. In fact limiting $0\leq k\leq0.1$, leads to an $R^2= 0.98$ for the concentration.
% \art{do we explain that L2 minimization can lead to better predictions }

 % using the first moment alone we cannot predict the radius, but it is possible to completely recover the radius with higher moments.

% Only by including the second and fourth moment in the kernel model for predicting the radius do we achieve models that have learning curve that converge. Thus all of our learning curves furthermore corroborate our previous conclusion: using the first moment alone we cannot predict the radius.

Finally, from Figure~\ref{fig:stabfreq}, we see that the learning curve saturates around
a maximum wavenumber of $0.8$. This indicates that $0<k<0.8$ is the ideal range to measure particles in the range $0<a<2$.
% Although to measure more details of the particles geometry, it is likely that $ka > 1.6$ will be required.

% \begin{figure}
%   \centering
% \begin{tikzpicture}
%   \node[inner sep=0pt] (0,0)
%     {\includegraphics[height=0.2\linewidth]{../Images/min_wavelength.pdf}};
%   \node at (-0.3,1.) {$\lambda$};
% \end{tikzpicture}
%  \caption{Compares smallest wavelength necessary, $\lambda = 2\pi/0.8$, to measure all particle radiuses. The particles shown have radiuses $0.2, 0.8, 1.4,$ and $2.0$.
% % Clearly multiple scattering encodes sub-wavelength information~\cite{simonetti_multiple_2006}.
% }
%  \label{fig:wavelength}
% \end{figure}

% In conclusion, these results corroborate our previous conclusion:
% with only the first moment we can recover the concentration but not the radius, and when using the first and second moment, we can recover both the concentration and the radius.
 \emph{\textbf{Conclusions {\textemdash}}} our results indicate that the first direct backscattered moment $\ensem{u}$ does not carry information about a broad range of particle radiuses, for strong scatterers. However, the second moment $\ensem{u}_2$ does carry this information. On the other hand, the particle concentration can be accurate predicted from just $\ensem{u}$. We also demonstrated that only incident wavenumbers $0<k<0.8$ are needed to accurately measure particles with radius $0<a<2$. This implies that neither theory, simulation or experiments need go beyond $ka = 1.6$, at least for strong scatterers. This also means that we are able to accurately recover radiuses that 20 times smaller than the smallest incident wavelength.

In this study we did not consider limitations in spatial and temporal resolution, which of course are important in practise. However, before specialising to one particular scenario, i.e. typical acoustics and light scattering experiments, we need to know what is possible to measure or not in an ideal setting. Studies like these are therefore a vital first step. {Another important step is to quantify how uncertainties in the measurements affect the prediction of the particle properties. This can be achieved by using Guassian process regression~\cite{rasmussen2006gaussian}, which is, in a sense, a Bayesian version of kernel ridge regression.}

{Ultimately, our machine learning model could be embedded into a device to predict particulate properties. Though our model is initially trained on simulated data, our training procedure is simple enough that the model can be updated using real data.
This step of adapting models trained on simulated data to real applications has been applied to challenging problems such as robotic grasping~\cite{simtorealrobo}, facial recognition~\cite{NIPS2017_6612}, 3D pose inference~\cite{3dpose} and optical flow estimation~\cite{MIFDB18} to name a few. These applications have advanced in strides by using simulated data, and we see a similar potential for characterising random media, such as this work.}
% This step of adapting models trained on simulated data to real applications\footnote{In particular in robotics, adapting a machine learning model trained on simulated data to real world applications is referred to as sim-to-real~\cite{simtorealrobo}} has already been applied to challenging physical problems such as
% robotic grasping~\cite{simtorealrobo}, facial recognition~\cite{NIPS2017_6612}, 3D pose inference~\cite{3dpose} and optical flow estimation~\cite{MIFDB18} to name a few. These applications have advanced in strides by using simulated data, and we see a similar potential for advancing traditional physics applications, on which this work takes the first step.   }

% Our predictions for both the radius and concentration were accurate, with $R^2 = 0.93$ and $R^2 = 0.96$, even though
% the standard error of the mean of the training and test set were $10\%$, see the tops graphs of Figure~\ref{fig:stabnummed}. The backscattered signal were also sampled coarsely in time, with only 29 points on each signal. These imprecisions are promising, because it indicates that it may be possible to accurately measure particle size and concentration using receivers with limited resolution and limited number of backscattering measurements.

Both the simulation (near-surface backscattering) and machine learning approach we have presented could be applied to characterise any type of particulate material from wave backscattering. To extend our approach, to 3D and other types of particles, computational efficiency is important. Simulating the backscattered moments would be faster if the multilevel Monte Carlo methods
\cite{giles_multilevel_2015} and fast multipole methods~\cite{zhang_fast_2007} were used. For instance, it may be possible to measure the physical properties of the particles, as well as the size and concentration. Another avenue to create more backscattering data is to piece together different theoretical models, which could then be validated with the numeric approach we introduced: near surface backscattering in time.

% \emph{Acknowledgments \textbf{{\textemdash}}}
A.L. Gower, W.J. Parnell and I.D. Abrahams are grateful for the funding provided by EPSRC (EP/M026205/1,EP/L018039/1). R.M. Gower is grateful for funding provided by the FSMP at the INRIA - SIERRA project-team.
J. Deakin would like to acknowledge the receipt of an EPSRC CASE studentship from the School of Mathematics and Thales UK.

\newpage

\begin{center}
\textbf{\large  Supplementary material on Characterising particulate random media from near-surface backscattering}
\end{center}
%%%%%%%%%% Merge with supplemental materials %%%%%%%%%%
%%%%%%%%%% Prefix a "S" to all equations, figures, tables and reset the counter %%%%%%%%%%
\setcounter{equation}{0}
\setcounter{figure}{0}
\setcounter{table}{0}
\setcounter{page}{1}
\makeatletter
\renewcommand{\theequation}{S\arabic{equation}}
\renewcommand{\thefigure}{S\arabic{figure}}
% \renewcommand{\bibnumfmt}[1]{[S#1]}
% \renewcommand{\citenumfont}[1]{S#1}
%%%%%%%%%% Prefix a "S" to all equations, figures, tables and reset the counter %%%%%%%%%%

Here we explain how to reproduce our results shown in the letter \emph{Characterising particulate random media from near-surface backscattering}, including high performance software \href{https://github.com/jondea/MultipleScattering.jl/tree/v0.1}{to simulate the backscattering}, \href{https://github.com/gowerrobert/MultipleScatteringLearnMoments.jl}{implement the machine learning} and how to access the \href{https://zenodo.org/record/1126642#.WusNMnXwb8s}{data} used.

\emph{\textbf{Calculating near-surface backscattering \textemdash}} We choose the multi-pole method because it easily accommodates circular particles, it is very accurate and it has hardly any artefacts\cite{martin_multiple_2006}. It has been the method of choice for other packages dedicated to multiple scattering~\cite{mackowski_multiple_2011}, and
 can be made computationally efficient with the fast multi-pole method\cite{zhang_fast_2007}. As this method is well established, here we only give a brief outline. Our code~\cite{art_gower_2017_1133989} was implemented in Julia~\cite{bezanson2017julia}, a language focused on high performance numerics, and is open source~\cite{jondea_multiplescattering.jl:_2017}. All the tests and benchmarks we refer to are reproduced in the \href{https://github.com/jondea/MultipleScattering.jl/tree/v0.1.1/example}{example} and \href{https://github.com/jondea/MultipleScattering.jl/tree/v0.1.1/test}{test} folder. The data used in the paper is also available online~\cite{gower_dirichlet_data_2018}.

The $j$-th particle scatters a wave $u^j_{\Out}$, which satisfies the 2D scalar wave equation $\nabla^2 u^j_{\Out} + k^2 u^j_{\Out} =0$, and therefore has the form
\be \label{eqn:outwaves}
	u^j_{\Out} = \sum_{m=-M}^M A^j_m J_m (k_\Out a) \frac{H_m(k_\Out r^j)}{H_m(k_\Out a)} \ee^{\ii m \theta^j} \;\;\; \text{for} \;\; r^j \geq a,
\en
where $a$ is the particle radius, $(r^{j} ,\theta^{j} )$ are the polar coordinates of $(x,y)$ centred at the $j$-th particles centre $(x_j,y_j)$ and $M$ is chosen so that~\eqref{eqn:outwaves} converges. The $H_m$ and $J_m$ are Hankel and Bessel functions of the first kind, and the $A^j_m$ are to be determined from boundary conditions. Using the above, we write the backscattered wave in the form $u_b = \sum_{j=1}^N u^j_{\Out}$,  where $N$ is the number of particles. The boundary conditions
\begin{equation}
  u = 0 \quad \text{on} \;\; r^j =a \;\; \text{for} \;\; j =1,\ldots, N,
\end{equation}
where $u = \ee^{\ii k ( x- x_R - t)} + u_b$, and Graf's addition theorem leads to
\begin{multline}
 \label{eqns:As}
	A^s_m  + \sum_{n=-M}^M \sum_{\stackrel{j =1}{j\not= s}}^N A^j_n \frac{J_n (k_\Out a)}{H_n(k_\Out a)}  H_{n-m}(k_\Out R^{js})\ee^{\ii (n-m) \theta^{js}}   \\ = -\ii^{m} \ee^{\ii k_\Out (x^s -x_R)},
\end{multline}
for $m=-M,\ldots,M$ and $s=1,\ldots, N$. We use the above to solve for the $A^s_m$ and completely determine $u_b$. The point $(x_s,y_s)$ is the centre of the $s$-th inclusion and $(R^{js},\theta^{js})$ are the polar coordinates of $(x_s,y_s)$ centred at $(x_j,y_j)$.

After calculating the solution in the frequency domain for $0\leq k \leq 1$, we can calculate the backscattered response in time measured at $(x_R,0)$, where we consider $a k \leq 2$. The smaller $ak$, the smaller $M$ needs to be.
 % Substituting~\eqref{eqns:As} into \eqref{eqn:outwaves} completely determines $u_b$.
\begin{figure}[ht!]
  \includegraphics[width=0.7\linewidth]{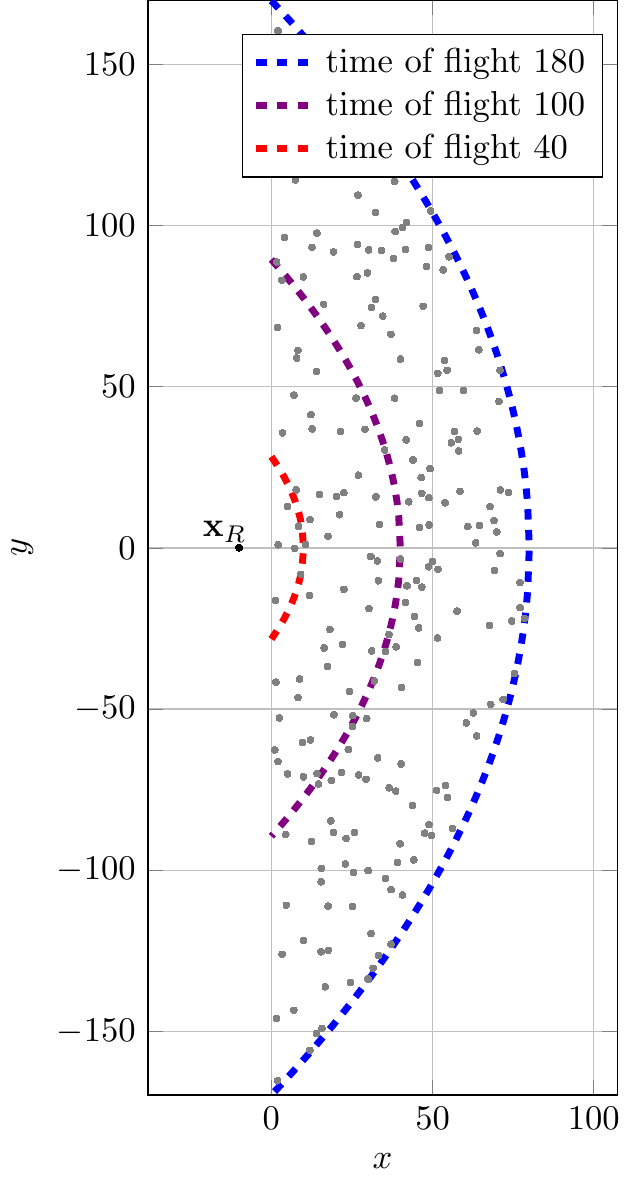}
  \caption{ shows particles randomly placed according to a uniform distribution. For a plane incident wave to travel from $x=-10$ to any point on the blue dashed curve and then directly back to the receiver $\mathbf x_R$ takes time $t=180$. Likewise for the purple/red curve it takes $t=120/60$. Note the phase speed of background is 1 (non-dimensional).}
  \label{fig:TimeOfFlight}
\end{figure}

\begin{figure}[ht!]
  \includegraphics[width=0.9\linewidth]{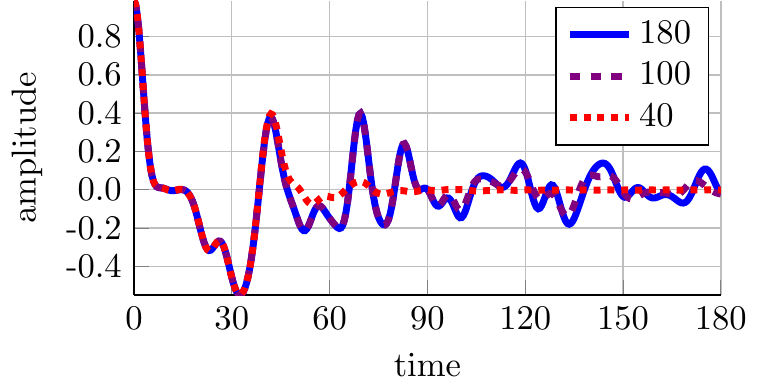}
  \caption{The backscattering of the incident wave $\ee^{-0.1(x-x_R - t)^2}$, received at $(x_R,0)$, from simulations where there are no particles that took longer than time {\color{red}{40}}, {\color{purple}{100}} and {\color{blue}{180}} for their first scattered wave to arrive at $(x_R,0)$. Figure~\ref{fig:TimeOfFlight} shows the configuration of these particles.}
  \label{fig:plot_TimeOfFlight}
\end{figure}

One notable challenge, is that we want to approximate the backscattering $u_b$ from a infinite halfspace $x>0$ filled with particles. One option is to use a computational domain large enough for the backscattered signal to converge
% . This issue of convergence is rarely discussed in the literature
\cite{galaz_experimental_2010,chekroun_time-domain_2012,pinfield_simulation_2012,muinonen_coherent_2012}. In our numerical experiments, on the order of $10^4$ particles are needed before the backscattering converges within 1\%. This becomes particularly challenging when $k a \approx 1$ or larger, because $M$ needs to increase.
We find a simple solution is to calculate the backscattering in time $t$ and keep only the early arrival $t<98$. That way we exclude contributions from particles further away from the surface, where it takes longer than $t=100$ for their first scattered wave to return to the receiver $\mathbf x_R = (x_R,0)$. This allows us to only simulate the response from particles near the surface.
 % y = sqrt((t - (x -lx))^2 - (x -lx)^2)
The backscattering from the particles to the left of the blue dashed line in Fig.~\ref{fig:TimeOfFlight} is shown by the blue curve in Figure~\ref{fig:plot_TimeOfFlight}, and likewise for the purple/red curve. All three backscatterings are the same up to time $t = 40$, as it takes $t>40$ for scattered wave from the closest particle above the red curve in Figure~\ref{fig:TimeOfFlight} to arrive at $\mathbf x_R$. The same rationale explains why the blue and purple curves are the same for $t < 100$.

For the backscattered signal to converge within 1\% accuracy required a 1800 mesh points for the wavenumber evenly sampled in $0\leq k \leq2$. This is because the finer the mesh, the longer the time period of the discrete Fourier transform of the backscattered waves. A long time period is necessary because, due to multiple scattering, the backscattering can last a long time. On the contrary, if the time period considered is too short, the discrete Fourier transform will no longer be causal~\cite{lyons_understanding_2010}.

% \section*{Learning from backscattering }
\emph{\textbf{Learning from backscattering \textemdash}} To train our models to predict the radius and concentration, we use $L$ simulated media where $ (r_\ell,v_\ell) \in \R^2$ is the particle radius and the concentration of the $\ell$-th media.  Let  $\ensem{u^{\ell}}_j$ be the $j$th centred moment of the simulated backscattered waves from the $\ell$th media, and let
\begin{equation}
  (\vec M^\ell):= \{ \ensem{u^{\ell}}_j \, | \, j=1, 2, \ldots, m \},
\end{equation}
be the collection of $m \in \mathbb{N}$ moments. We will refer to   $(\vec M^\ell , r_\ell, v_\ell )$ as the \emph{training set} throughout.
 For the results presented in the article we used only the mean backscattering $\ensem{u}_1 = \ensem{u}$, $m=1$, or the mean and second moment, $m=1,2$, of the backscattering.

% \subsection{Kernel ridge regression}

\emph{\textbf{Kernel ridge regression \textemdash}} Our objective is to train $h^r: \vec M^\ell \rightarrow h^r(\vec M^\ell) \in \R_+ $ and  $h^v:\vec M^\ell \rightarrow h^r(\vec M^\ell) \in \R_+ $  to predict the radius and concentration, respectively.
In kernel ridge regression, these have a parametric form
\begin{equation}
  h^r(\vec M) = \sum_{\ell =1} \alpha_{\ell}^r K( \vec M^\ell,\vec M),
\end{equation}
and
\begin{equation}
  h^v(\vec M) = \sum_{\ell =1} \alpha_{\ell}^v K( \vec M^\ell,\vec M),
\end{equation}
where
 $K: (\vec M', \vec M ) \rightarrow \R$ is a given kernel function, $\alpha_{\ell}^r$, and $\alpha_{\ell}^v$ for $\ell =1,\ldots, L$ are the parameters that need to be determined. Let ${\bf K} :=  \left(K(\vec M^n,\vec M^\ell)\right)_{n \ell}$ be the kernel matrix,  $\vec \alpha := (\alpha_{ \ell})_{\ell=1}^L$, $ \vec r := (r_\ell)_{\ell=1}^L$ and $\vec v := (v_\ell)_{\ell=1}^L.$ We calculate the unknown parameter vectors $ \vec \alpha$ by minimizing the $L^2$ loss over the training set. That is, to determine the parameters $\alpha^r_{\ell}$ of the $h^r$ model we solve
%$h^r(\vec M^\ell)$ is close to $r_\ell$ and $h^v(\vec M^\ell)$ is close to $v_{\ell}$ for all $\ell =1,\ldots, L$.
\begin{align}\label{eq:ac90n11d}
\vec \alpha^r &=  \arg\min_{ \vec \alpha\in \R^{L}}\frac{1}{2L} || {\bf K} \vec \alpha-\vec r||_2^2 + \frac{\lambda_r}{2} \langle {\bf K} \vec \alpha, \vec \alpha \rangle,
\end{align}
where $\lambda_r >0$ is the regularization parameter.
Analogously we introduce a regularization parameter $\lambda_v$ for the $h^v$ model.
% Using the solution to the above we can recover our model by using that
% \begin{equation}
%   h^a(\vec M) = \langle \vec w_*^a, \psi(\vec M )\rangle \overset{\eqref{eq:Kernalclass}}{ =}  \sum_{\ell =1} \alpha_{* \ell}^a K( \vec M^\ell,\vec M) = \vec K \vec \alpha_*^a.
% \end{equation}

The kernels we tested in our experiments are the following.
\begin{equation}
\begin{array}{l@{\qquad}l}
\mbox{Gaussian}&K(\vec M,\vec M') = \exp\left(\frac{-||\vec M-\vec M'||^2}{2\sigma^2}\right)\\
\mbox{Ornstein--Uhlenbeck}&K(\vec M,\vec M') = \exp\left(\frac{-||\vec M-\vec M'||}{\sigma}\right) \\
\mbox{Linear}&K(\vec M,\vec M') = \Tr{\vec M^\top \vec M'},
\end{array} \label{eq:kernels}
\end{equation}
where $\sigma>0$ is the kernel parameter.

%\subsection{Selecting the moments using feature selection}
%Before fitting a model, we determined which moments we should incorporate in the model.

\begin{figure}
 	\centering
 	 \includegraphics[width =  0.33\textwidth]{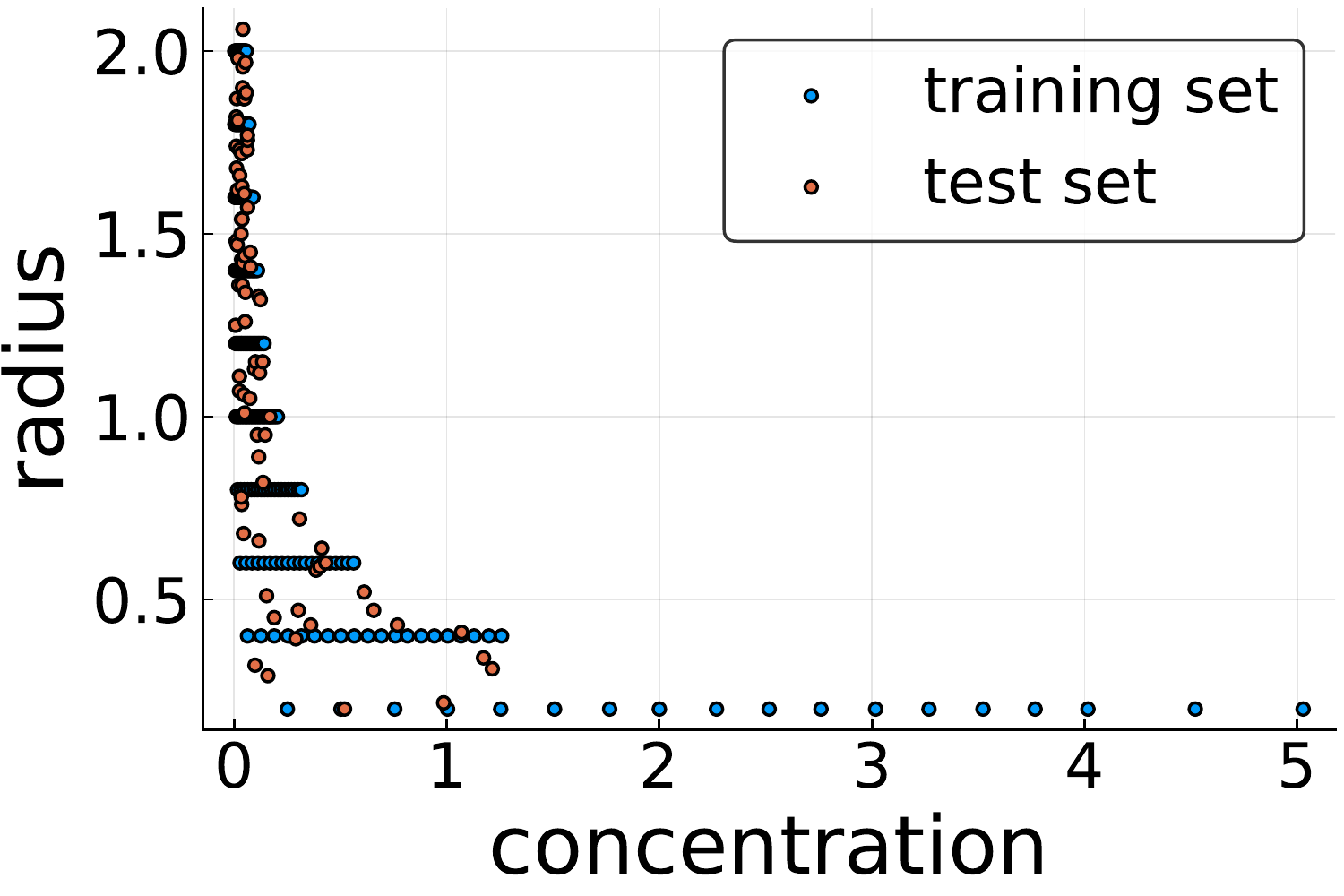}
 \includegraphics[width =  0.33\textwidth]{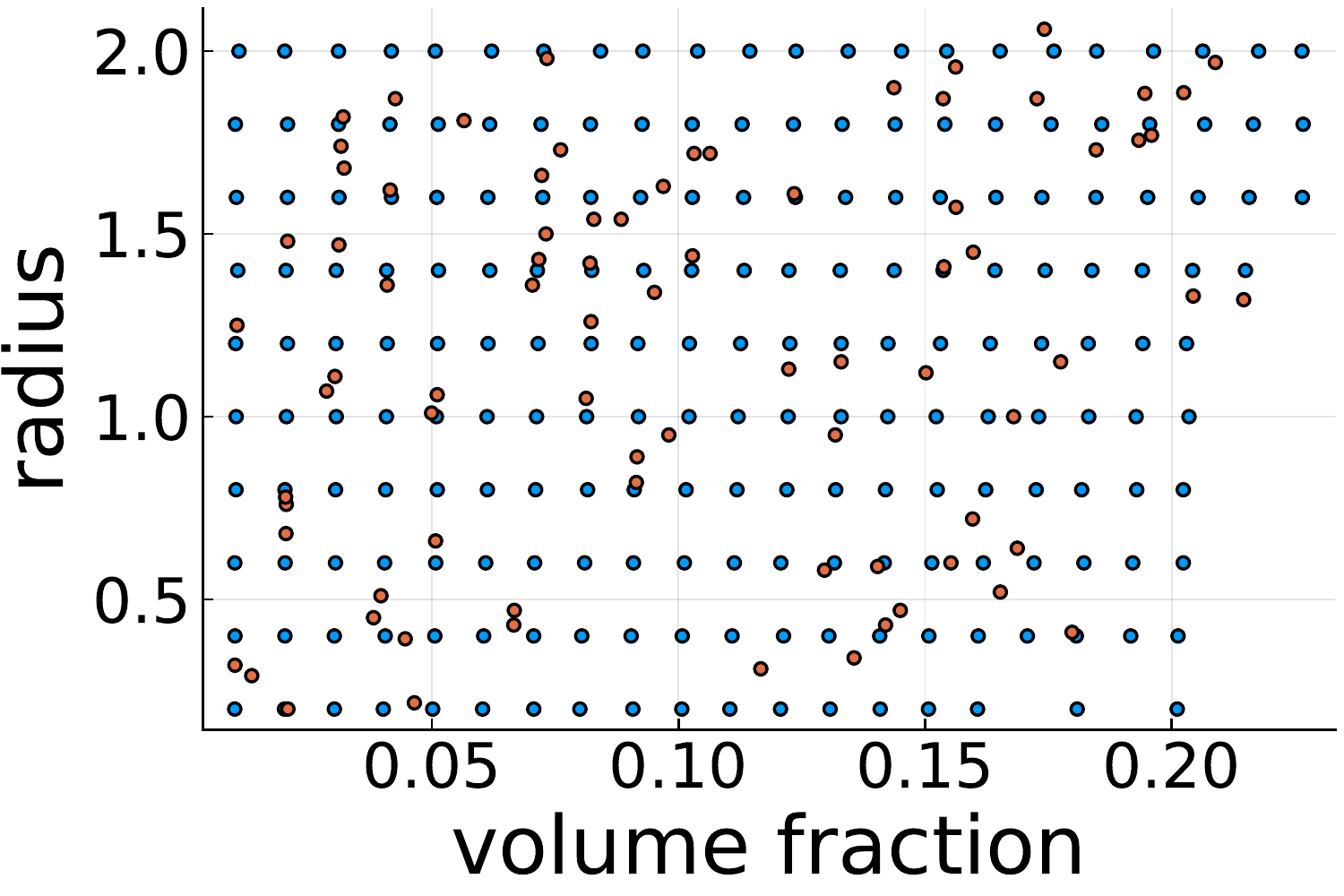}
 \caption{On the top (bottom) we have the concentration and radius (volume fraction and radius) of each medium in the training set (in blue) and test set (in orange).}
 \label{fig:trainvstest}
\end{figure}

 % \subsection{Testing the model}
 \emph{\textbf{Testing the models \textemdash}} To validate our models, we produced a test set of media with randomly chosen particle radiuses and concentrations, none of which are part of the training set. Let $(\vec M^t, r_t,v_t) $ for $t= 1, \ldots, T$ be this test set, see Figure~\ref{fig:trainvstest} for a scatter plot comparing the training set and test set.

 To measure the goodness of fit of our models, we use $R^2$, the R squared coefficient of determination, over the test set. For example, let $\hat{r}_t = h^r(\vec M^t)$
 be the predicted radiuses for $t = 1,\ldots, T,$ and let $\bar{r} = \sum_{t=1}^T r_t$ be the average radius over the test, then
 \begin{equation}
   R^2 = 1 - \frac{\sum_{t=1}^T (\hat{r}_t - r_t)}{\sum_{t=1}^T ( \bar{r} -r_t )}.
 \end{equation}
 If $R^2$ is close to $1$, then the $\hat{r}_t$ are significantly better at predicting the true radiuses in comparison to using the mean $\bar{r}$ as the predicted radius. Otherwise, if $R^2$ is close to zero or even negative, then the predicted $\hat{r}_t$ are worse than using the mean $\bar{r}$.
% Bare in mind that the mean over the test is a conservative base line, since we do know the mean over the test set in practice and thus cannot build this base line model. This is why

% \subsection{Implementation details}

\emph{\textbf{Implementation details \textemdash}} The code for the kernel ridge regression~\cite{gower_multiplescatteringlearnmoments:_2017} based on moments was implemented in the Julia programming language, where we also show
\href{https://github.com/gowerrobert/MultipleScatteringLearnMoments.jl/tree/master/examples/moments_data}{how to calculate the moments} from the full simulated data.
% and can be readily retrieved from~\url{https://github.com/gowerrobert/MultipleScatteringLearnMoments}.
 The parameters $\lambda_r$, $\lambda_v$ and $\sigma$ were all determined using a $7$-way cross-validation over the training set. No parameters were hand picked.

We also carried out standard data pre-processing including, normalizing and centring the data. We also applied the natural logarithm to the radius and concentration in the training set. Thus to recover a predicted radius and concentration, we apply exponentiation. This explicitly enforces that the models predicts a positive radius and concentration.

% This includes the regularization parameters and {\color{red}eventual} kernel parameters.
\emph{\textbf{Acknowledgments \textemdash}} A.L. Gower, W.J. Parnell and I.D. Abrahams are grateful for the funding provided by EPSRC (EP/M026205/1,EP/L018039/1). R.M. Gower is grateful for funding provided by the FSMP at the INRIA - SIERRA project-team.
J. Deakin would like to acknowledge the receipt of an EPSRC CASE studentship from the School of Mathematics and Thales UK.

\bibliography{MultipleScattering.bib,SharedMultipleScattering.bib}% Produces the bibliography via BibTeX.

\end{document}